\documentclass[aps,showpacs,nofootinbib,twocolumn]{revtex4}
\usepackage[english]{babel}
\usepackage{amsfonts,amsmath,amssymb,mathrsfs,epsf}
\usepackage{hyperref}
\usepackage[mathcal]{euscript}
\newcommand{\eq}{\begin{equation}}
\newcommand{\feq}{\end{equation}}
\newcommand\lsim{\mathrel{\rlap{\lower4pt\hbox{\hskip1pt$\sim$}}
\raise1pt\hbox{$<$}}}
\newcommand\gsim{\mathrel{\rlap{\lower4pt\hbox{\hskip1pt$\sim$}}
\raise1pt\hbox{$>$}}}
\begin{document}
\bibliographystyle{prsty}
\title{The Effect of Inhomogeneities on the Luminosity
Distance--Redshift Relation:\\
is Dark Energy Necessary in a Perturbed Universe? }
\author{Enrico Barausse}\email{barausse@sissa.it}
\affiliation{Dipartimento di Fisica ``G.\ Galilei,'' Universit\`{a}
di Padova, via Marzolo 8, Padova
I-35131, Italy\\ and \\
SISSA/ISAS, via Beirut 4, I-34014 Trieste, Italy}
\author{Sabino Matarrese}\email{sabino.matarrese@pd.infn.it}
\affiliation{Dipartimento di Fisica ``G.\ Galilei,'' Universit\`{a}
di Padova, and INFN, Sezione di Padova, via Marzolo 8, Padova
I-35131, Italy}
\author{Antonio Riotto}\email{antonio.riotto@pd.infn.it}
\affiliation{INFN, Sezione di Padova, via Marzolo 8, I-35131, Italy}
\date{\today}
\begin{abstract}
\noindent The luminosity distance--redshift relation is one of the
fundamental tools of modern cosmology. We compute the luminosity
distance--redshift relation in a perturbed flat matter-dominated
Universe, taking into account the presence of  cosmological
inhomogeneities up to second order in perturbation theory.
Cosmological  observations implementing the luminosity
distance--redshift relation tell us  that the Universe is
presently undergoing a phase of accelerated expansion. This seems
to call for a mysterious Dark Energy component with negative
pressure. Our findings suggest  that the need of a Dark Energy
fluid may be challenged once a  realistic  inhomogeneous Universe
is considered and that an accelerated expansion may  be consistent
with a matter-dominated Universe.
\end{abstract}
 \maketitle
\section{Introduction}
\noindent
One of the fundamental relations in cosmology is the
one expressing  the luminosity distance $d_L$  of a cosmological
source in terms of  its redshift $z$.  In recent years
this relation has been exploited to measure
the present value of the expansion rate,
Hubble's constant, with increasing accuracy.
With the exploration of the Universe at redshifts of order unity,
we now have information about the time evolution of the expansion rate
\cite{riess}.  A most surprising result is that the time evolution of the
expansion rate does not seem to be described by a matter-dominated
Friedmann-Robertson-Walker (FRW) homogeneous
cosmological model of the Universe.  The usual
explanation for the discrepancy is that there is a new component of the energy
density of the Universe, known as Dark Energy, that determines the recent
evolution of the expansion rate.  Of course all indications for Dark Energy are
indirect; from cosmological data we only  infer that
the Universe is presently undergoing a phase
of acceleration.

Since the luminosity distance--redshift relation
 is of such fundamental importance, we
must understand any possible effects that would result in a
relation different from the FRW prediction for a homogeneous
Universe.
 The computation of the
luminosity  (or angular diameter)
distance in a locally non homogeneous universe was
first addressed  by Zeldovich \cite{zel}). Dyer and  Roeder \cite{dr} adopted
the so-called empty beam approximation to derive an equation for the angular
diameter distance. Our techinique instead relies on the traditional
perturbative approach, recently used in Refs.
\cite{sasakidist,sasakifoll,pynedist}.

In this paper we study the
change in the luminosity distance--redshift relation due to
cosmological perturbations present in a background matter-dominated Universe
with zero cosmological constant. In
particular, we provide the generic expression for the luminosity
distance--redshift relation at second order in perturbation
theory, for any value of the physical redshift and any
direction of observation. We only consider modifications to the
luminosity distance--redshift relation  of a matter-dominated
Universe, although our results can be easily extended to a
Universe containing a mixture of matter and other fluids. Our
findings may be used, for instance,  to estimate the influence of
lensing on the brightness of supernovae sources.
What our second-order procedure does not  account for
is another   class of terms,
where a first-order small deflection in the light ray leads the
geodesic to deflect to a region where the perturbations are sizeable
relative to the neighbourhood of the geodesic in the unperturbed
space-time.

It is well-known that the luminosity distance--redshift relation
allows to extract the theoretical predictions for the local Hubble
rate ${\cal H}$ and deceleration parameter $q$ upon expanding
around the observer point at $z=0$. In a perturbed Universe,
however, both the Hubble constant and the deceleration parameter
lose  their deterministic nature; one has to consider the
statistical nature of the vacuum fluctuations from which the
present-day gravitational potential is originated. Therefore, the
theoretical predictions for the expected values of the
cosmological parameters are accompanied by a nonvanishing cosmic
variance implying an intrinsic theoretical error. When calculating
the variance of the deceleration parameter, we uncover an
interesting infrared effect. We observe, at second order in
perturbation theory, a large contribution to the variance from the
cosmological perturbations with the largest wavelengths. If
inflation is the origin of the cosmological perturbations
\cite{lr}, the beginning of inflation provides a cut-off to the
infrared modes of the fluctuations. Our results suggest that if
the super-Hubble modes  have physical significance, they could
yield a most important modification to the deceleration parameter.
One might even speculate that a complete treatment of the effect
could   obviate the need for the Dark Energy assumption.  Indeed,
despite the fact that the homogeneous FRW model for the
matter-dominated Universe and no cosmological constant
predicts that the latter may not
decelerate, {\it i.e.} $q_0=\frac{1}{2}>0$, because of the large
variance, the true locally-determined value of the deceleration
parameter has a non-zero probability of being less than
zero. Put it differently, the theoretical prediction in  a
perturbed matter-dominated model is not a single well-defined
curve in the luminosity-distance redshift plane, but instead is
represented by a finite confidence region whose size is determined
by the cosmic variance.

The paper is organized as follows.
In the next section we obtain the generic luminosity distance--redshift
relation  valid at any order in perturbation theory around the homogeneous
FRW background. In Section III we express the luminosity distance--redshift
relation in terms of the metric perturbations expanded up to second order.
Section IV is devoted to the evaluation of the mean and the variance
of the cosmological observables and to the discussion of their implications.
Finally, Section IV present our conclusions and comments.
All the technical details are included in  Appendices A and  B.

\section{The generic luminosity distance--redshift relation}
\noindent
The goal of this section is to provide the reader with the
generic formulation of the luminosity distance--redshift relation in a
general setting. Our treatment closely follows the one given in Ref.
\cite{sasakidist}. In the next section, we will make use of  this
generic formulation to obtain the luminosity distance--redshift relation
in a perturbed FRW Universe being the  background matter-dominated Universe
with zero cosmological constant.

In order to deal with the propagation of light
from a given source to the observer, we make use of
 the conformal (Weyl) invariance of the electromagnetic field and define
\begin{equation} \label{a1}
d\hat{s}^2=\hat{g}_{\mu\nu}dx^{\mu}dx^{\nu}
=a^2\left(\eta\right)g_{\mu\nu}dx^{\mu}dx^{\nu}\;,
\end{equation}
where $x^{\mu}=\left(\eta,x^i\right)$ are the space-time
coordinates and the scale factor $a$ is normalized to unity at the
present conformal time $\eta_0$ ($a(\eta_0)=a_0=1$).

We will use a ``$\hat{\phantom{g}}$'' to mark quantities
calculated in a space-time with the metric $\hat{g}_{\mu\nu}$;
quantities without ``$\hat{\phantom{g}}$'' will be calculated with
the metric $g_{\mu\nu}$ instead. Furthermore, quantities
with a ``$\widetilde{\phantom{g}}$'' stand for quantities in a
perturbed Universe.

Using the geometric optics approximation, the energy-momentum
tensor of a photon emitted  by a given source is
\begin{equation} \label{a2}
\hat{T}^{\mu\nu}=A^2\hat{k}^{\mu}\hat{k}^{\nu}\;,
\end{equation}
where $A$ is the scalar amplitude of the wave and $\hat{k}^{\mu}$
is the photon four-momentum:
\begin{eqnarray} \label{a3}
\hat{k}^{\mu}&=&\frac{dx^{\mu}}{dv}=\hat{g}^{\mu\nu}\partial_{\nu}S,
\qquad \hat{k}^{\mu}\hat{k}_{\mu}=0,\nonumber \\
\hat{k}^{\nu}\hat{\nabla}_{\nu}\hat{k}^{\mu}&=&\frac{d^2x^{\mu}}{dv^2}+
\hat{\Gamma}^{\mu}_{\alpha\beta}\frac{dx^{\alpha}}{dv}
\frac{dx^{\beta}}{dv}=0\;.
\end{eqnarray}
Here  $S$ is the phase (eikonal) of the wave and $v$ is an affine
parameter along the ray ({\it i.e.} along the photon trajectory).
The evolution of the
parameter $A$ is  provided by the continuity equation
$\hat{\nabla}_\nu \hat{T}^{\mu\nu}=0$: we have
\begin{equation} \label{a5}
\frac{dA}{dv}=-\frac{1}{2}A\hat{\theta}\;,\qquad
\hat{\theta}=\hat{\nabla}_{\mu}\hat{k}^{\mu}\;.
\end{equation}
All these equations can be replaced by the corresponding ones in
the metric $g_{\mu\nu}$. If we define
\begin{equation} \label{a6}
d\lambda=a^{-2}dv\;, \qquad
k^{\mu}=\frac{dx^{\mu}}{d\lambda}=a^2\hat{k}^{\mu}\;,
\end{equation}
we easily get
\begin{eqnarray} \label{a7}
k_{\mu}&=&\partial_{\mu}S\;, \quad k^{\mu}k_{\mu}=0\;,\nonumber \\
k^{\nu}\nabla_{\nu}k^{\mu}&=&\frac{d^2x^{\mu}}{d\lambda^2}+
\Gamma^{\mu}_{\alpha\beta}
\frac{dx^{\alpha}}{d\lambda}\frac{dx^{\beta}}{d\lambda}=0\;,\nonumber\\
\frac{d\left(Aa\right)}{d\lambda}&=&-\frac{1}{2}Aa\theta\;,\quad
\theta=\nabla_{\mu}k^{\mu}\;.
\end{eqnarray}
To obtain the equation
describing the transport of the optical scalar $\theta$
(\textit{expansion} of the null congruence $k^\mu$) along the ray,
it is useful to recall the concept of null (Newman-Penrose)
tetrad: it is made of two real ($k^{\mu}$ and
$m^{\mu}$) and two complex conjugate ($t^{\mu}$ and $\bar{t}^{\mu}$)
4-vectors, satisfying
\begin{eqnarray} \label{a10}
k^{\mu}m_{\mu}&=&\bar{t}^{\mu}t_{\mu}=1\;,\nonumber \\
k^{\mu}k_{\mu}&=&
m^{\mu}m_{\mu}=t^{\mu}t_{\mu}=k^{\mu}t_{\mu}=m^{\mu}t_{\mu}=0\;.
\end{eqnarray}
Defining the projector
\begin{equation} \label{a11}
{\cal P}^{\mu}_{\nu}\equiv
t^{\mu}\bar{t}_{\nu}+
t_{\nu}\bar{t}^{\mu}=
\delta^{\mu}_{\nu}-k_{\nu}m^{\mu}-m_{\nu}k^{\mu}\;,\end{equation}
one can easily obtain the decomposition (the semicolon indicates
covariant differentiation in the conformal space-time with metric
$g_{\mu\nu}$)
\begin{equation} \label{a12}
k_{\mu;\nu}=a_{\nu}k_{\mu}+b_{\mu}k_{\nu}+A_{\mu\nu}\;,
\end{equation}
with
\begin{eqnarray} \label{a13}
A_{\mu\nu}&=&k_{\alpha;\beta}{\cal P}^{\alpha}_{\mu}{\cal P}^{\beta}_{\nu}
=2\Re\left(\frac{\theta}{2} t_{\mu}\bar{t}_{\nu}+\sigma
t_{\mu}t_{\nu}\right)\, ,\nonumber\\
 \label{a14}
\quad a_{\mu}k^{\mu}&=&b_{\mu}k^{\mu}=0\;,
\end{eqnarray}
where $\theta$ is real and  $\sigma$ is complex.
With such a decomposition we obtain
\begin{equation} \label{a15}
\theta=\nabla_{\mu}k^{\mu}\;,\,\,\,\,
\vert\sigma\vert^2=\frac{1}{2}
\left[k_{\mu;\nu}
k^{\mu;\nu}-\frac{\theta^2}{2}\right]\, .
\end{equation}
The quantity $\sigma$ is called  \textit{shear}
of the null congruence.

>From the Ricci identity
\begin{equation} \label{a16}
k_{\mu;\nu\delta}-\left(\nu\leftrightarrow \delta\right)
=R_{\sigma\mu\nu\delta}k^{\sigma}\;,
\end{equation}
contracting the indices $\mu$ and $\nu$, we obtain (see for instance
\cite{SEF})
\begin{equation} \label{a17}
\frac{d\theta}{d\lambda}=-
R_{\mu\nu}k^{\mu}k^{\nu}-\frac{\theta^2}{2}-2\vert\sigma\vert^2\;.
\end{equation}
In a similar fashion, the transport equation for the shear
$\sigma$ can be derived by contracting (\ref{a16}) with
$\bar{t}^{\mu}\bar{t}^{\nu}k^{\delta}$ and assuming that
$\bar{t}^{\mu}$ is parallel-transported along the ray; we get
\begin{equation} \label{a18}
\frac{d\sigma}{d\lambda}=-\sigma\theta+C_{\alpha\beta\nu\mu}
k^{\alpha}k^{\mu}\bar{t}^{\nu}\bar{t}^{\beta},\quad
k^{\nu}\nabla_{\nu}\bar{t}^{\mu}=0\;,
\end{equation}
where $C_{\alpha\beta\nu\mu}$ is the Weyl tensor.

To calculate the luminosity distance, we need the energy flux
per unit surface, $\ell$,  measured by an observer with
4-velocity $\hat{u}^{\mu}$
\begin{equation} \label{a19}
\ell=\sqrt{\hat{h}_{\mu\nu}
\left(\hat{u}_{\sigma}\hat{T}^{\sigma\mu}\right)
\left(\hat{u}_{\delta}\hat{T}^{\delta\nu}\right)}=A^2\omega^2\;,
\end{equation}
where
\begin{equation} \label{a20}
\hat{h}_{\mu\nu}=\hat{g}_{\mu\nu}+\hat{u}_{\mu}\hat{u}_{\nu}\;,\qquad
\omega=-\hat{u}_{\mu}\hat{k}^{\mu}\;.
\end{equation}
If the source has  physical radius $R$ (which we will eventually set
to zero) and if we choose the affine parameter $\lambda$ on
the ray connecting the observer and the source such that
$\lambda=0$, $\lambda=\lambda_s$,
$\lambda=\lambda_s+\Delta\lambda_s$ correspond to the observer
today, to the surface of the source and to its center, the
power ${\cal L}$ emitted by the source and  measured by a comoving observer,
is
\begin{equation} \label{a21}
{\cal L}=4 \pi R^2\ell\left(\lambda_s\right)\;.
\end{equation}
The luminosity distance thus reads
\begin{equation} \label{a22}
d_L=R
\sqrt{\frac{\ell\left(\lambda_s\right)}{\ell\left(0\right)}}=
R
\frac{A\left(\lambda_s\right)}{A\left(0\right)}
\left(1+\widetilde{z}\left(\lambda_s\right)\right)\;,
\end{equation}
where  the redshift $\widetilde{z}$ is defined by
\begin{equation} \label{a23}
1+\widetilde{z}\left(\lambda_s\right)
=\frac{\omega\left(\lambda_s\right)}{\omega\left(0\right)}\;.
\end{equation}
To summarize, the evaluation of the luminosity distance goes
through the following steps: one has to solve for the photon
trajectory  from which
one may deduce the frequency $\omega$; then one is able to solve
for the expansion parameter
 $\theta$ and the shear $\sigma$; finally one solves for the amplitude
$A$.

\section{The luminosity distance--redshift relation in a perturbed FRW
Universe}
The formalism we have summarized  in the previous Section can now
be applied to obtain the luminosity
distance--redshift relation in a perturbed FRW
Universe. In particular, we are interested in a perturbative
treatment up to second order. This Section contains only the main
steps of the calculation, a plethora of details can be found in
Appendix A.

First of all, we need to expand the wave four-vector of the photon
reaching the observer at second order, \eq\label{defk}
k^\mu=k_{(0)}^\mu+k_{(1)}^\mu+k_{(2)}^\mu\;.\feq Similarly, we
expand the photon trajectory  as
\begin{equation}\label{a33}
x^{\mu}(\lambda)=x_{(0)}^{\mu}(\lambda)+
x_{(1)}^{\mu}(\lambda)+x_{(2)}^{\mu}(\lambda)\;.
\end{equation}
Both quantities  $k_{(r)}^\mu$ and the $x_{(r)}^\mu$ ($r=0,1,2$)
are fully determined by the geodesic equation
\begin{eqnarray}\label{a34}
&&\frac{d}{d\lambda}\left(k_{(0)}^{\mu}+k_{(1)}^{\mu}+
k_{(2)}^{\mu}\right)
=-\Gamma^\mu_{\alpha\beta}k^\alpha
k^\beta\nonumber\\
&=&-\Gamma^{\mu}_{(1)\alpha\beta}k_{(0)}^{\alpha}k_{(0)}^{\beta}
-\Big(\Gamma^{\mu}_{(2)\alpha\beta}k_{(0)}^{\alpha}k_{(0)}^{\beta}\nonumber
\\&+&2\Gamma^{\mu}_{(1)\alpha\beta}k_{(0)}^{\alpha}k_{(1)}^{\beta}
+\Gamma^{\mu}_{(1)\alpha\beta,\gamma}
k_{(0)}^{\alpha}k_{(0)}^{\beta}x_{(1)}^{\gamma}\Big)\:,
\end{eqnarray}
and by the initial conditions (given in Appendix A).
%More precisely, it is useful to define
%\begin{equation} \label{a24}
%K^{\mu}=-\frac{k^{\mu}}{\omega\left(0\right)a
%\left(\eta\left(0\right)\right)}\;,
%\end{equation}
%so  to have
%\begin{equation} \label{a25}
%g_{\mu\nu}u^{\mu}K^{\nu}(\lambda=0)=1\;.
%\end{equation}
%From now on, we will therefore mean $k^{\mu}=K^{\mu}$ and
%$\lambda$ such that $\frac{dx^{\mu}}{d\lambda}=k^{\mu}=K^{\mu}$
%(let's note that now we have $\lambda_s>0$, $\Delta\lambda_s>0$):
%of course, equations (\ref{a7}), (\ref{a8}), (\ref{a17}),
%(\ref{a18}) keep unchanged under this affine transformation.\\
We then need to perturb the expansion $\theta$ of our
null congruence \eq
\theta=\theta_{(0)}+\theta_{(1)}+\theta_{(2)}\;:\feq these
functions are  determined by Eq. (\ref{a17}) and by the boundary
conditions. Furthermore, since the  right-hand side of Eq.
(\ref{a17}) contains the square of the shear, we need also to
solve Eq. (\ref{a18}) up to first order, again with suitable
boundary conditions. After the quantities $\theta$, $\sigma$ and
$A$ are known, one is ready to calculate the luminosity
distance--redshift relation. Sending the physical size $R$ of the
source (and therefore $\Delta\lambda_s$) to zero, we get the
following  expression

\begin{eqnarray}\label{pp}
d_L&=&\left(1+\widetilde{z}\left(\lambda_s\right)\right)a_0\lambda_s
\Bigg[1-\frac{1}{2}\int_{0}^{\lambda_s}\theta_{(1)}
\left(\lambda,\lambda_s\right)d\lambda\nonumber\\&+&\frac{1}{8}
\left(\int_{0}^{\lambda_s}\theta_{(1)}
\left(\lambda,\lambda_s\right)d\lambda\right)^2\nonumber\\
&-&\frac{1}{2}\int_{0}^{\lambda_s}
\theta_{(2)}\left(\lambda,\lambda_s\right)d\lambda
-k_{(1)}^0\left(\lambda_s\right)\nonumber\\&-&k_{(2)}^0
\left(\lambda_s\right)+\frac{1}{2}k_{(1)}^0\left(\lambda_s\right)
\int_{0}^{\lambda_s}\theta_{(1)}\left(\lambda,\lambda_s\right)d\lambda\Bigg].
\end{eqnarray}
This is the generic expression for the luminosity
distance--redshift relation at second order in perturbation theory
and for any value of the physical redshift
$\widetilde{z}(\lambda_s)$ and any given direction of
observation.

Since the
physical observable is the redshift and not the
affine parameter $\lambda_s$, we have to trade  $\lambda_s$ in terms
of $\widetilde{z}$ in Eq. (\ref{pp}). In order to achieve this
we may make use of the definition of the redshift in the unperturbed
Universe

\begin{equation}
1+z=\frac{a_0}{a(\eta_0-\lambda_s)}\,, \,\,\,
\frac{\lambda_{s}}{\eta_0}=1-\frac{1}{\sqrt{1+z}}\,,
\end{equation}
from which we deduce the relation between the physical redshift
$\widetilde{z}$ and $z$

\begin{align}\label{a700}
&1+\widetilde{z}\left(\lambda_s\right)=\frac{a_0}
{a\left(\eta\left(\lambda_s\right)\right)}
\frac{\left(u_{\mu}k^{\mu}\right)\left(\lambda_s\right)}{\left(u_{\mu}
k^{\mu}\right)\left(0\right)}\nonumber\\
&=a_0\left(1-\left(k_{(1)}^0\left(\lambda_s\right)+k_{(2)}^0
\left(\lambda_s\right)\right)\right)\nonumber\\
&\times\Bigg[ a\left(\eta_{(0)}\left(\lambda_s\right)\right)
+a'\left(\eta_{(0)}\left(\lambda_s\right)\right)\left(\eta_{(1)}
\left(\lambda_s\right)+\eta_{(2)}\left(\lambda_s\right)\right)\nonumber\\
&+\frac{1}{2}a''\left(\eta_{(0)}\left(\lambda_s\right)\right)
\eta_{(1)}\left(\lambda_s\right)^2\Bigg]^{-1}\nonumber\\
&=(1+z)\left(1+T_1(\lambda_s)+T_2(\lambda_s)\right)\;,
\end{align}
where $\eta_{(1)}$ and $\eta_{(2)}$ are respectively the
first- and second-order expansion of the photon conformal time (see
Appendix A) and
\begin{eqnarray}\label{t1}
T_1\left(\lambda\right)&=&\left(-\frac{a'(\eta_{(0)})}{a(\eta_{(0)})}
\eta_{(1)}-k_{(1)}^0\right)\left(\lambda\right),\\
\label{t2} T_2\left(\lambda\right)&=&
\Bigg(-\frac{a'(\eta_{(0)})}{a(\eta_{(0)})}
\eta_{(2)}-k_{(2)}^0-\frac{a''(\eta_{(0)})}{2a(\eta_{(0)})}
\eta_{(1)}^2\nonumber\\&+&\left(\frac{a'(\eta_{(0)})}{a(\eta_{(0)})}
\eta_{(1)}\right)^2
+\frac{a'(\eta_{(0)})}{a(\eta_{(0)})}\eta_{(1)}k_{(1)}^0
\Bigg)(\lambda).\end{eqnarray}  We may therefore compute first the
luminosity distance--redshift relation in terms of the parameter
$z$ and then as a function of the physical redshift
$\widetilde{z}$ using Eqs. (\ref{a700}), (\ref{t1}) and (\ref{t2})
to express $z$ as a function of $\widetilde{z}$. Yet, in the
Appendix we describe another procedure, which allows to obtain an
explicit expression of $d_L$ as a function of the physical
redshift $\widetilde{z}$. Both give rise to the same result.

To compute all these quantities we have to
expand the metric to second order. We adopt  the  comoving
and synchronous gauge\footnote{The   luminosity distance--redshift
relation is a gauge-invariant concept as explicitly shown in Ref.
\cite{sasakidist} and therefore the choice of the gauge is
arbitrary.}
\begin{eqnarray}
\label{mmm}
ds^2&=&-d\eta^2+\gamma_{ij}dx^i dx^j\, ,\nonumber\\
\gamma_{ij}&=&
(1-2\psi_{(1)}-\psi_{(2)})\delta_{ij}+
\chi_{(1)ij}+\frac{1}{2}\chi_{(2)ij}\;,\end{eqnarray}
where both $\chi_{(1)ij}$ and $\chi_{(2)ij}$ are traceless.

Assuming that the source and the observer are both comoving with the
fluid, one has (at any order in perturbation theory)
$u^{\mu}=\delta^{\mu}_{0}$.
By solving Einstein's equations, the metric  perturbations
$\psi_{(r)}$ $\chi_{(r)}$, $r=1,2$, can be expressed in terms
of the peculiar gravitational potential $\varphi$ \cite{desitter}:
\begin{eqnarray}\label{chi1}
\chi_{(1)ij}&=&-\frac{\eta^2}{3}\left(\varphi_{,ij}-
\frac{1}{3}\delta_{ij}\nabla^2\varphi\right)\;,
\\\label{phi1}
\psi_{(1)}&=&\frac{5}{3}\varphi+\frac{\eta^2}{18}\nabla^2\varphi\;,
\end{eqnarray}
at first order and
\begin{equation}
\label{phi2} \psi_{(2)}=-\frac{50}9
\varphi^2-\frac{5\eta^2}{54}
\varphi^{,k}\varphi_{,k}+
\frac{\eta^4}{252}
\left(-\frac{10}{3}
\varphi^{,ik}\varphi_{,ik}+\left(\nabla^2\varphi\right)^2\right)\;,
\end{equation}
\begin{multline}\label{chi2}
\chi_{ij}^{(2)} =\frac{\eta^4}{126}\Big(19\varphi^{,k}_{~~,i}
\varphi_{,kj}-12 \varphi_{,ij} \nabla^2\varphi +4
(\nabla^2\varphi)^2 \delta_{ij}\\
-\frac{19}{3}\varphi^{,kl}\varphi_{,kl}
\delta_{ij}\Big)-\frac{10\eta^2}{9}\left(\varphi_{,i}\varphi_{,j}-\frac13
\varphi^{,k}\varphi_{,k}\delta_{ij}\right)+\pi_{ij}\;,
\end{multline}
at second order\footnote{The well-known residual gauge ambiguity
of the synchronous and comoving gauge has been fixed as in Ref.
\cite{desitter}.}. We have disregarded both linear vector modes
(since they are not generated during inflation) and linear tensor
modes (because their dynamical role is negligible).  At
second order tensor modes described by  $\pi_{ij}$ do not enter in
the computation of the Hubble rate and the deceleration
parameter.
Furthermore, the values of the second-order potentials have been computed
with a proper match to
the initial conditions set by  single-field models of inflation
\cite{en}.\footnote{Notice, however, that the initial conditions do not
play a significant role, since physical observables like the Hubble constant
and the deceleration parameter depend only upon the
rate of change of cosmological inhomogeneities.}

In Appendix A we will give explicit expressions for
$k_{(r)}^\mu(\lambda)$ and  $\theta_{(r)}$, $r=1,2$ in terms of
$\varphi$: in particular we will see that the luminosity distance
$d_L$, given by Eq. (\ref{a70}), involves, besides the physical
redshift $\widetilde{z}$  and the direction of
observation, only the peculiar gravitational potential $\varphi$
and its gradients along the background geodesic connecting the
source to the observer and the present conformal time
$\eta_0=2/{\cal H}_0$.
\section{Results and implications}
\noindent We are now ready to extract from the findings of the
previous Section the relevant cosmological parameters in a
realistic perturbed Universe. To this purpose, let us recall that
in a generic unperturbed FRW Universe the following relation
holds
\begin{equation}\label{q0obsdef}
d_L=\frac{c}{{\cal H}_0}
\left[z+\frac{z^2}{2}
\left(1-q_0\right)+{\cal O}\left(z^3\right)
\right]\;,
\end{equation}
where ${\cal H}_0$ is the unperturbed Hubble constant and $q_0$ is
the unperturbed deceleration parameter. Therefore, an observer
measuring a relation of the type $d_L= {\cal A}+ {\cal B}
\widetilde{z}+{\cal C} \widetilde{z}^2+\cdots$ would conclude that
${\cal B}$ and ${\cal C}$ provide a measure of the Hubble constant
and deceleration parameter at the present epoch. To make contact
with this procedure, we may expand the generic formula
(\ref{pp})
for the luminosity distance (valid at any value of the
redshift $\widetilde{z}$) around $\widetilde{z}=0$. In this way,
we may determine the value of the Hubble constant and the
deceleration parameter that would be measured by an observer
having  at her/his  disposal sources with redshifts
$\widetilde{z}\lsim 1$ in a {\it perturbed} Universe. Performing
properly the angular average $\langle \cdots \rangle_\Omega$ and
comparing this expansion of Eq. (\ref{pp}) with Eq.
(\ref{q0obsdef}), we infer  the  expression for  $\widetilde{{\cal
H}}_0$ and for $\widetilde{q}_0$ (see Appendix B for details):
\begin{widetext}
\begin{eqnarray}
\label{H} \langle\widetilde{{\cal H}}_0\rangle_\Omega&=&{\cal
H}_{0} \Bigg[1-\frac{2}{{\cal
H}_0}\left(\frac12\psi_{(1)}^\prime+\frac14\psi_{(2)}^\prime
+\psi_{(1)}\psi_{(1)}^\prime+\frac{1}{12}
\left(\chi_{(1)}^{ij}\right)^\prime\chi_{(1)ij}\right)\Bigg]\nonumber\\
&=&{\cal H}_{0}\bigg[1-\left(\frac{1}{18}\nabla^2\varphi-
\frac{5}{108}\left(\nabla\varphi\right)^2+
\frac{5}{27}\varphi\nabla^2\varphi\right)\left(\frac{2}{{\cal
H}_0}\right)^2
-\left(\frac{1}{189}\varphi^{,ij}\varphi_{,ij}+\frac{1}{252}
\left(\nabla^2\varphi\right)^2\right)\left(\frac{2}{{\cal H}_0}\right)^4\,\bigg]\;,\\
\label{q}
\langle\widetilde{q}_0\rangle_\Omega&=&\frac{1}{2}\Bigg[1
+\frac{2}{{\cal H}_0}\Bigg(2\psi_{(1)}^\prime+\psi_{(2)}^\prime+
4\psi_{(1)}\psi_{(1)}^\prime+
\frac{1}{3}\left(\chi_{(1)}^{ij}\right)^{\prime}\chi_{(1)ij}\Bigg)
\nonumber\\
&+&\left(\frac{2}{{\cal
H}_0}\right)^2\Bigg(\frac12\psi_{(1)}^{\prime\prime}
+\frac14\psi_{(2)}^{\prime\prime}+\frac94\left(\psi_{(1)}^\prime\right)^2
+\psi_{(1)}\psi_{(1)}^{\prime\prime}+
\frac{7}{40}\left(\chi_{(1)ij}\right)^\prime\left(\chi_{(1)}^{ij}\right)^\prime
+\frac1{12}\left(\chi_{(1)}^{ij}\right)^{\prime\prime}\chi_{(1)ij}\Bigg)
\nonumber\\& +&\left(\frac{2}{{\cal H}_0}\right)^3
\Bigg(\frac12\psi_{(1)}^\prime\psi_{(1)}^{\prime\prime}
+\frac{1}{60}\left(\chi_{(1)ij}\right)^\prime
\left(\chi_{(1)}^{ij}\right)^{\prime\prime}\Bigg)
\Bigg]\nonumber\\&=&\frac{1}{2}\bigg[1+
\left(\frac{5}{18}\nabla^2\varphi+
\frac{25}{27}\varphi\nabla^2\varphi-
\frac{25}{108}\left(\nabla\varphi\right)^2\right)\left(\frac{2}{{\cal
H}_0}\right)^2+
\left(\frac{1}{30}\left(\nabla^2\varphi\right)^2+\frac{23}{270}\,
\varphi^{,ij}\varphi_{,ij}\right)\left(\frac{2}{{\cal
H}_0}\right)^4 \bigg]\;,\end{eqnarray}
\end{widetext}
where $\psi_{(r)}$, $\chi_{(r)}$ ($r=1,2$) and $\varphi$ are
evaluated at the observer's position $\boldsymbol{x}_{(0)}=0$ and
at the present time $\eta_0$.\footnote{Note that Eq. (\ref{H})
agrees with Eq. (39) of \cite{kmnr}.} Eqs. (\ref{H}) and (\ref{q})
are among the main results of our paper. The key point is that the
values (\ref{H}) and (\ref{q}) of the Hubble constant and
 the deceleration parameter in a perturbed
Universe are not  deterministic. Indeed, we must consider the statistical
nature of the vacuum fluctuations from which the present-day linear
gravitational potential $\varphi$ is originated. This implies that
the gravitational potential does not have well-defined values, but one can only
define  the probability of finding a given value at a given point in space.
In a given realization of the perturbed Universe, the values
of the cosmological parameters may change
in regions of the Universe which are causally disconnected today. Therefore,
it  is unavoidable that the theoretical predictions for the values of the
cosmological parameters come with a nonvanishing cosmic variance which
implies an intrinsic theoretical error.

It will turn out  that the variance
 can be large and get its largest contribution from the infrared
super-Hubble modes of the metric perturbations.

Let us   evaluate the expected uncertainty in the determination of
the deceleration parameter, we treat the metric fluctuation
$\varphi$ as a Gaussian random variable with zero statistical
mean over a volume which, in the inflationary picture,
has a size much larger than the present-day Hubble
radius. The variable $\varphi$    takes random
values over different ``realizations''. We may express the variance
in terms of the matter power spectrum.  The procedure would be to
fix a spherical domain of large volume ${\cal V}$ surrounding the
observer and containing the most distant sources of
interest\footnote{ We are interested in comparing the theoretical
predictions of a perturbed Universe to the ones obtained in a
unperturbed FRW homogeneous classical background. Since the
quantities relative to the unperturbed model are recovered in the
zero-momentum limit, the volume average has to be performed over a
volume of size at least as large as the present-day Hubble
radius.}. However, since the main contribution to the variance
comes from the longest wavelengths, this volume averaging is
irrelevant\footnote{ As a technical remark, we stress that the
low-pass filtering procedure over the volume ${\cal V}$ should
have been performed before Taylor expanding the luminosity
distance--redshift relation (\ref{pp}) around the observer point
at $\widetilde{z}=0$. This amounts to cutting off the spurious
contribution of ultraviolet modes to the means and variances of
physical observables which would appear performing the average
after expanding in powers of the physical redshift.}. We will
express $\varphi$ and its derivatives in terms of a Fourier
integral, so
\begin{eqnarray}
\varphi &=& \int\frac{d^3\!k}{(2\pi)^3}
\ \varphi_{\vec{k}} \ e^{i\vec{k}\cdot\vec{x}} , \quad
\varphi_{,i} = \int\frac{d^3\!k}{(2\pi)^3} \ ik_i
\ \varphi_{\vec{k}} \ e^{i\vec{k}\cdot\vec{x}}\, ,\nonumber\\
\nabla^2\varphi &=& - \int\frac{d^3\!k}{(2\pi)^3} \ k^2
\ \varphi_{\vec{k}} \ e^{i\vec{k}\cdot\vec{x}} , \quad {\rm etc.}
\end{eqnarray}
The Fourier components $\varphi_{\vec{k}}$ satisfy
\begin{eqnarray}
\label{fourpoint}
\overline{\varphi_{\vec k}}  =  0\, , \,\,
\overline{\varphi_{\vec{k}_1} \varphi_{\vec{k}_2}}
&=&  (2 \pi)^3 \delta^{(3)}(\vec{k_1}+\vec{k_2}) \
 P_{\varphi}(k_1)\\
\overline{\varphi_{\vec{k}_1}\varphi_{\vec{k}_2}
\varphi_{\vec{k}_3} \varphi_{\vec{k}_4}}
 &=&  (2\pi)^6
\left\{\delta^{(3)}(\vec{k}_1+\vec{k}_2) \delta^{(3)}(\vec{k}_3+\vec{k}_4)
\right.
\nonumber\\
&\times& \left.\ P_{\varphi}(k_1)P_{\varphi}(k_3)
 +({\rm cyclic}\,\,{\rm  terms}) \right\}\, , \nonumber
\end{eqnarray}
where $\overline{(\cdots)}$ denotes the statistical average and
$P_\varphi(k)$ is the power spectrum of the gravitational
potential.  We can express $P_\varphi(k)$ in terms of the
matter power spectrum as
\begin{equation}
P_{\varphi}(k) \equiv
\frac{9\pi^2}{2} \ a_0^4 {\cal H}_0^4 \ \frac{\Delta^2(k,a_0)}
{k^7},
\label{pphipdelta}
\end{equation}
where $\Delta^2(k,a_0)$ is the (dimensionless) power spectrum of the
matter density fluctuations linearly extrapolated to the present
time. We express the power spectrum
$\Delta^2(k,a_0)$ in terms of the transfer function $T^2(k)$.  For a
Harrison--Zel'dovich spectrum, the power spectrum is
\begin{equation}
\Delta^2(k,a_0) = {\cal A}^2 \left(\frac{k}{a_0{\cal H}_0}\right)^4 T^2(k) ,
\label{powerspectrum}
\end{equation}
where ${\cal A}$ is the dimensionless amplitude, ${\cal A}=1.9\times10^{-5}$.

An analysis similar to the one performed in Ref. \cite{kmnr} shows that
the biggest contribution to the variance of the deceleration parameter
comes from the terms of the type $\varphi\nabla^2\varphi$ whose variance
is
\begin{eqnarray}
\textrm{Var}\left[\varphi\nabla^2\varphi \right] &\simeq&
\left( \frac{9}{4}  a_0^4 {\cal H}_0^4 \right)^2\,
\int\frac{dk_1}{k_1}\Delta^2(k_1,a_0)\nonumber\\
&\times&  \int\frac{dk_2}{k_2^5}
\Delta^2(k_2,a_0)\nonumber\\
&\simeq & \textrm{Var}\left[\varphi \right]\,
\textrm{Var}\left[\nabla^2\varphi \right]  \,.
\end{eqnarray}
Taking into account that $T(k)\rightarrow 1$ when $k\rightarrow 0$, we conclude
that the variance of the deceleration parameter is sensitive also to
the infrared modes of the Harrison--Zel'dovich
power spectrum and is
\begin{equation}
\frac{\sqrt{\textrm{Var}\left[\langle\widetilde{q}\rangle_\Omega\right]}}{q_0}
\simeq {\cal A}^2\,\ln\frac{k_\textrm{MAX}}{k_\textrm{MIN}}\sim
10^{-10}\,\ln\frac{k_\textrm{MAX}}{k_\textrm{MIN}},
\label{logterm2}
\end{equation}
where $k_\textrm{MIN}$ is the infrared cut-off and
$k_\textrm{MAX}$ is the ultraviolet cut-off set by the averaging
volume. We can take it to coincide with the horizon volume, that
is $k_{\rm MAX}={\cal H}_0^{-1}$.

Rather than a Harrison--Zel'dovich spectrum, if we assume a slightly
red spectrum so that $\Delta^2(k)\propto k^{3+n_s}$ with $0 <
(1-n_s) \ll 1$, then the logarithmic term in
Eq.\ (\ref{logterm2}) is replaced by $(1-n_s)^{-1}
(k_{\rm MAX}/k_\textrm{MIN})^{(1-n_s)}$. Now this will give unit variance if

\begin{equation}
\ln
k_{\rm MAX}/k_\textrm{MIN}\simeq(45+\ln (1-n_s))/(1-n_s)\, .
\end{equation}

The variance of the
deceleration parameter is infrared sensitive. There are various
ways to avoid such a sensitivity:
\begin{itemize}

\item  If the power spectrum is blue, {\it i.e.}
$n_s>1$, the variance does not pick up a large contribution from the
long wavelengths;

\item One may imagine to fix the physical infrared cut-off at a momentum
scale corresponding to the present-day Hubble radius, but in the
inflationary picture there is no strong motivation to do so since
inflation is likely to have lasted for a period much longer than
the minimum required number of {\it e}-folds  $\sim 60$. After
inflation, the size of the Universe which is relatively
homogeneous is of the order of $\ell_*\sim H_{\rm I}\,e^N$, where
$N$ is the total number of {\it e}-folds and $H_{\rm I}$ is the
Hubble rate during inflation. As the length scale $\ell_*$ is much
larger than the present-day Hubble size, it seems reasonable to
sum up  all the super-Hubble modes up to a length scale
corresponding to
 $\ell_*$ at the end of inflation\footnote{One should also keep in
mind that
on scales larger than $\ell_*$ the Universe may become extremely inhomogeneous
due to quantum fluctuations produced during inflation and a considerable
part of the physical volume of the entire Universe may remain forever in the
inflationary phase \cite{eternal}.}.

\item The presence of an infrared sensitivity
might be  only an artifact of the perturbative expansion and it might
disappear when dealing with a complete nonperturbative approach. To
check this possibility, a computation beyond second order
is needed, even though we do not see any a priori reason why
such cancellation should manifest itself at higher orders.

\end{itemize}

Since our findings reveal  the crucial role
played by the infrared modes, a deeper understanding of their
physical significance is certainly desirable\footnote{For references
dealing with the physical significance of the super-Hubble modes
{\it neglecting} the gradients see Refs.
\cite{listir} and \cite{k2}. Recently, it is has been claimed \cite{noh} that
``the relativistic zero-pressure fluid perturbed to second order in a flat
Friedmann background coincides with the Newtonian result'' and that
``there are no relativistic correction terms even near and beyond the horizon
to the second-order perturbation''. However, this claim relies on
disregarding second-order tensor modes and is not justified,
as implied by the discussion of Ref. (\cite{silent}), where
it was stressed the importance of second-order tensor modes for the correct
recovery of the Newtonian limit from the full relativistic theory.}.
A large  variance of the deceleration parameter
$\frac{\sqrt{\textrm{Var}\left[\langle\widetilde{q}\rangle_\Omega
\right]}}{q_0}$ is caused  by  the fact that
the cosmological perturbations on super-Hubble scales are time-dependent
when gradients are consistently taken into account.
Indeed, it is easy to convince oneself that, were the
infrared part of the perturbations
$\psi_{(r)}$, $\chi_{(r)}$ ($r=1,2$) constant in time, they might be
eliminated from the metric (\ref{mmm}) by a simple rescaling of the
spatial coordinates which allows to remain in the synchronous gauge.
Such a freedom is lost if the inhomogeneities have a non-trivial
time-dependence on super-Hubble scales. Consider, for instance, the
first-order perturbation
$\psi_{(1)}=\frac{5}{3}\varphi+\frac{\eta^2}{18}\nabla^2\varphi$. It
contributes to
 the metric (\ref{mmm}) by  a piece $\left(1-\frac{10}{3}\varphi
\right)\delta_{ij}$ which can be rescaled to $\delta_{ij}$ by a
transformation of the spatial coordinates. However, this change is not for
free,  it gives rise to the crucial  piece proportional
to $\varphi\nabla^2\varphi$ at second order as it is easily
realized by inspecting the transformations of the
perturbations listed in Ref. \cite{desitter}.
The fact that the variance gets its largest contribution from
a piece proportional to $\varphi\nabla^2\varphi$ does not
then come as a surprise, it manifests the impossibility of
rescaling out the super-Hubble modes. An alternative way of dealing with
super-Hubble modes can be found in Appendic C.

In the following, we set the infrared cut-off to the value fixed by
the beginning of inflation which, in turn, depends upon the total number
of $e$-folds $N$ of the inflationary period.

The ratio $(k_{\rm MAX}/k_\textrm{MIN})$ is predicted by the inflationary
theory to be (we are adopting natural units)

\begin{equation}
\frac{k_{\rm MAX}}{k_\textrm{MIN}}\simeq 10^{-30}\left(\frac{T_{\rm RH}}
{H_{\rm I}}\right)\,e^N\, ,
\end{equation}
where $T_{\rm RH}$ is the reheating temperature at the beginning of the
radiation era after the end of inflation.

For the variance to be of the order of the background value
$q_0=\frac{1}{2}$,
the perturbation spectrum
would have to extend to a factor of $\exp(6\times10^{18})$ ($10^{18.8}$
$e$-folds!) times the present Hubble radius
(for $T_{\rm RH}\simeq H_{\rm I}$) in the case of the Harrison--Zel'dovich
spectrum. However, if,
for instance,  $n_s=0.94$ on
super-Hubble-radius scales, then a variance of order unity is obtained if the
perturbation spectrum extends $N\sim 700$
 $e$-folds beyond the Hubble radius. Since the
present Hubble radius corresponds to a scale that crossed the Hubble radius
about 60 $e$-folds before the end of inflation, if inflation lasted more than
$\sim 700$  $e$-folds with a super-Hubble-radius spectral index of $n_s=0.94$,
then the
effect of super-Hubble-radius perturbations on the
locally observed value of the deceleration parameter
would be sizeable.

What are the practical implications of our findings? What observations
tell us is that the Universe is presently undergoing a phase
of accelerated expansion, {\it i.e.} that the deceleration parameter
is negative. Indeed,
the unexpected faintness of high-redshift Type Ia supernovae (SNe Ia),
as measured by two independent teams \cite{riess},
has been interpreted as evidence that the expansion of the Universe
is accelerating. In an {\it unperturbed} FRW Universe, the deceleration
parameter is uniquely determined by the relative densities of
the various fluids with their own equation of state

\begin{equation}
q_0=\frac{1}{2}\Omega_0+\frac{3}{2}\sum_i\,w_i\,\Omega_i\, ,
\end{equation}
where $\Omega_0$ is the present-day total energy density
parameter and  $\Omega_i$ are the relative contributions of the various
components with equation of state $w_i=P_i/\rho_i$ ($P$ and $\rho$
are the pressure and energy density, respectively). Therefore,
the observation of a  negative value of the deceleration parameter
seems to call for the presence of a ``Dark Energy'' component with negative
equation of state \cite{rp} and abundance $\Omega_\Lambda\sim 0.7$.

The need of a mysterious Dark Energy fluid seems to be challenged once a
realistic {\it perturbed}
Universe is considered. Our results show that the theoretical predictions
for the local Hubble rate and the deceleration parameter are affected by
a cosmic variance whose size may be large, depending upon the
value of the spectral index and the overall duration of inflation.
Suppose, as we have done so far, that the Universe is globally flat
and matter-dominated with $\Omega_M\sim 0.3$ and  the {\it background} value
$q_0=\frac{1}{2}>0$.
Because of the large variance, however, the true locally-determined
value of the deceleration parameter has  non-zero probability
of being less than zero! In other words, in a perturbed Universe, acceleration
might not imply the existence of  Dark Energy. Fig. 1
gives a qualitative sketch of the effect of the cosmic variance
predicted in our paper as far as the magnitude--redshift
relation is concerned: the theoretical prediction in  a matter-dominated
model is not a single well-defined curve, but instead is
represented by a finite region whose size is determined by the
cosmic variance.
\begin{figure}
\epsfxsize=3.45in
\begin{center}
\leavevmode \epsffile{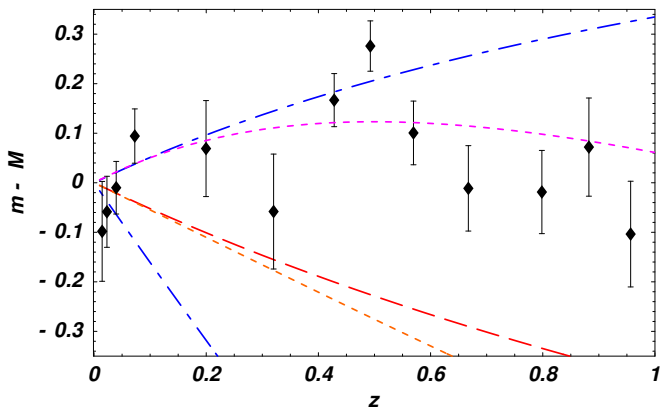}
\end{center}
\caption{The residual Hubble diagram with respect to an empty Universe
for a collection of type Ia Supernovae taken from Ref. \cite{riessbis}.
The apparent magnitude $m$ and the absolute magnitude $M$ are related to
the luminosity distance by the relation $(m-M)=-5+5\,{\rm log}_{10}\,
d_L({\rm pc})$.
The red long-dashed line corresponds to $q_0=1/2$ for a flat
matter-dominated Universe (up to ${\cal O}(z^3)$ terms); the blue
dot-dashed lines correspond to the values $q=q_0\pm1$, {\it i.e.} to the
possible size of the cosmic variance band expected in a perturbed
Universe; the orange short-dashed line corresponds to the prediction of the
flat matter-dominated
unperturbed Universe and the short-dashed purple line to a flat
unperturbed Universe
with
$\Omega_\Lambda=0.7$ and $\Omega_M=0.3$.}
\end{figure}

\section{Conclusions and comments}

Let us close  with some comments.  First of all,
let us  notice that in a perturbed Universe
the theory is unable to provide the
expected  value of the Hubble constant  (\ref{H}) and deceleration
parameter (\ref{q})
since   they
depend upon the ``bare'' unobservable Hubble constant
${\cal H}_0$. Therefore, in order
to compute the probability that a typical observer measures values
of the Hubble constant and deceleration parameter in agreement with the
observations, a complete  procedure would require marginalizing over
the bare Hubble constant ${\cal H}_0$ with some physical
prior. However, predicting a large
variance  for the deceleration parameter
can be already regarded as an indication that the tree-level value
$q_0=\frac{1}{2}$ may not be  in conflict with observing
a locally accelerating Universe. A  large variance
does not imply the breakdown of perturbation theory either. Indeed,
the density contrast $\delta\rho/\rho\sim (\varphi\nabla^2\varphi/a^2
{\cal H}^2)$
(valid in the synchronous gauge for {\it all} scales) is at most
of order unity. We have checked that a
 nonperturbative approach -- along the lines of Ref.
\cite{k2} -- leads to similar conclusions and that $\varphi\nabla^2\varphi$
is in fact the first term of the expansion of $e^{-\varphi}\nabla^2\varphi$
\cite{k3}. To get a physical intuition
of the reason why there is  a large contribution
to the variance of the deceleration parameter one might think in terms
of the energy density contrast. Indeed, a large variance
of the $\varphi\nabla^2\varphi$ term implies a large
 variance of $\delta\rho/\rho$ when summing up every Fourier
mode of the perturbations. Such a variance is dominated by the
infrared part of the spectrum and consequently is seen by an
observer restricted to a region of size comparable to  the present-day Hubble
radius as a classical energy density background. Such
interpretation is however not entirely correct because we are
dealing with variances and not with averages of the physical
observables. When inspecting the left-hand side of Einstein
equations, one may regard the contribution from the first- and
second-order gravitational potentials  as kinetic energy of the
gravitational field. Bringing them to the right-hand side and upon
averaging, one may think of such terms as contributing to the
classical energy density. However, this reasoning does not directly apply
to variances.

Instead of performing an angular average over the solid angle, we
might have determined the expected values and variances of
the Hubble rate and the deceleration
parameter
as a function of the direction of observation.
However, since the variance is dominated by the
infrared part of the spectrum, we expect that the local
anisotropies of the variance are as small as
$(\nabla^2\varphi/a^2 {\cal H}^2)$.

Finally, one should extend our computation to values of the
physical redshift larger than unity. Indeed, another, albeit
indirect, evidence of the presence of a Dark Energy fluid, is the
measurement of the location of the first Doppler peak of the CMB
anisotropies. Since the theoretical prediction for such a location
depends, in an unperturbed Universe, on the total energy density
parameter $\Omega_0$, from its measurement one may infer that
$\Omega_0$ is very close to unity. Since Dark Matter amounts
to only 30\% of the critical energy density,  it is concluded that
a consistent fraction of the energy density of the Universe is
made of Dark Energy. However, this deduction holds in an
unperturbed Universe only. On the contrary, one should compute the
location of the first Doppler peak in a real perturbed Universe.
This amounts to computing the angular diameter distance $d_A$
subtended by the sound horizon on the last scattering surface.
However, since the angular diameter distance is related to the
luminosity distance by the {\it reciprocity relation} $d_A=d_L/(1+z)^2$
at any order in perturbation theory, it suffices to determine the
luminosity distance. The expression (\ref{pp}) for the luminosity
distance--redshift relation holds for any redshift and any
direction of observation and is therefore suitable to perform
such a task. Furthermore, extending our computation (for instance
to higher
redshift) is required to compare the theoretical predictions of a
real perturbed Universe to other physical observables supporting
the Dark Energy picture in the unperturbed cosmology, such as
the
Integrated Sachs-Wolfe (ISW) effect and the transition from the
decelerating to the accelerating    phase at redshifts
of order
unity \cite{k3}. In this respect we notice that the contribution to the
variance of physical observables, as
$\frac{\sqrt{\textrm{Var}\left[\langle\widetilde{q}\rangle_\Omega\right]}}
{q_0}$, {\it increases} with time and therefore both the
ISW effect and the transition from the
decelerating to the accelerating    phase might well be
consistent with a perturbed flat matter-dominated Universe.
We also point out that the generic
expression (\ref{pp}) may be used to estimate the influence of
lensing on the brightness of supernovae sources. Finally, let us
reiterate once more that a deeper understanding of the physical
significance of long wavelength perturbations is certainly needed.

\begin{acknowledgments}
It is a pleasure to thank
N.  Afshordi, C. Baccigalupi, N. Bartolo, S. Bashinsky,
R. Blandford, E. Branchini,  S. Dodelson,
G. F. Giudice, R. Kolb, J. Lesgourgues, S. Lilly, A. Linde, E. Linder,
A. Notari, L.
Pilo,  C. Porciani, U. Seljak, N. Straumann and
S. White
for many enlightening conversations on the various
aspects of this work.
\end{acknowledgments}
\begin{widetext}
\appendix
\section{The luminosity distance - redshift relation}
\subsection{The geodesics equation} To perform the calculation, it
is useful to renormalize the photon four-momentum is such a way
that
\begin{equation} \label{a25bis}
g_{\mu\nu}u^{\mu}k^{\nu}(\lambda=0)=1\;.
\end{equation}
This is achieved by shifting  the original photon momentum by a
factor $-1/\left[\omega(0)a(\eta(0)) \right]$, where
$\eta(0)$ is the value of the background conformal time at the observer point.
. From now on, we
will work with the shifted photon momentum.

Expanding $k^\mu=k_{(0)}^\mu+k_{(1)}^\mu+k_{(2)}^\mu$, Eq.
(\ref{a25bis}) and the null normalization condition for $k^\mu$
(the second of Eq. (\ref{a7})) imply the following initial
conditions:
\begin{equation} \label{a32}
k_{(0)}^{\mu}=\left(-1,e^i\right),\quad \delta_{ij}e^ie^j=1,\quad
k_{(1)}^{0}\left(0\right)=k_{(2)}^{0}\left(0\right)=0\,.
\end{equation}
Expressing then $\Gamma^{\mu}_{(r)\alpha\beta}$, $r=1,2$ in terms
of the perturbations of the metric, equation (\ref{a34}) gives
\begin{equation}\label{a40}
\frac{dk^0_{(1)}}{d\lambda}=
\psi_{(1)}^\prime-\frac12\left(\chi_{(1)ij}\right)^\prime
e^ie^j=\frac{1}{3}\varphi_{,ij}e^ie^j\left(\eta_{0}-\lambda\right)
=\frac{1}{3}\left(\eta_{0}-\lambda\right)\frac{d^2\varphi}{d\lambda^2}\;,
\end{equation}
\begin{align}
&\frac{dk^i_{(1)}}{d\lambda}=-\partial^i\psi_{(1)}+2
e^i\left(e^j\partial_j\psi_{(1)}\right)-2e^i\left(\psi_{(1)}\right)^\prime
-\partial_k\chi^i_{(1)j}e^ke^j+e^k\left(\chi^i_{(1)k}\right)^\prime
+\frac12\partial^i\chi_{(1)kj}e^ke^j=-\frac{5}{3}\varphi^{,i}+\frac{10}{3}\varphi_{,j}e^je^i\nonumber\\&
+\frac{\left(\eta_{0}-\lambda\right)^2}{6}\varphi^{,i}_{\phantom{i},jk}e^je^k-\frac{2}{3}(\eta_0-\lambda)e_j\varphi^{,ij}=
-\frac{5}{3}\varphi^{,i}+\frac{10}{3}e^i\frac{d\varphi}{d\lambda}
+\frac{\left(\eta_{0}-\lambda\right)^2}{6}\frac{d^2\left(\varphi^{,i}\right)}{d\lambda^2}-\frac{2}{3}(\eta_0-\lambda)\frac{d\varphi^{,i}}{d\lambda}\;,
\end{align}
\begin{align}\label{a46}
&\frac{dk^0_{(2)}}{d\lambda}=\frac12\psi_{(2)}^\prime-\frac14\left(\chi_{(2)ij}\right)^\prime
e^ie^j+2\psi_{(1)}^\prime e^i
k_{(1)i}-\left(\chi_{(1)ij}\right)^\prime e_i
k_{(1)j}+\partial_\mu\psi_{(1)}^\prime
x_{(1)}^\mu-\frac12\partial_\mu\left(\chi_{(1)ij}\right)^\prime
e^ie^j x_{(1)}^\mu=\nonumber\\&
-\frac{5}{18}\left(\eta_{0}-\lambda\right)\varphi^{,i}\varphi_{,i}+\frac{1}{42}\left(\eta_{0}-\lambda\right)^3
\left(\varphi^{,ij}\varphi_{,ij}-\left(\nabla^2\varphi\right)^2\right)
-\frac{19}{126}\left(\eta_{0}-\lambda\right)^3\varphi^{,i}_{\phantom{i},j}\varphi_{,ik}e^ke^j
+\frac{5}{9}\left(\eta_{0}-\lambda\right)\left(\varphi_{,j}e^j\right)^2\nonumber\\&
+\frac{2}{21}\left(\eta_{0}-\lambda\right)^3\nabla^2\varphi\varphi_{,ij}e^ie^j
%-\frac{1}{21}\left(\eta_{0}-\lambda\right)^3S^{jk}e_je_k-\frac{2}{3}\left(\eta_{0}
%-\lambda\right)\tau^{jk}e_je_k
-\frac14\partial_\eta\pi_{ij}e^ie^j+\frac{2}{3}\left(\eta_{0}-\lambda\right)\varphi_{,ij}e^ik_{(1)}^j
+\frac{1}{3}e^ke^j\varphi_{,kj}\eta_{(1)}+\frac{1}{3}e^ke^j\varphi_{,kji}x_{(1)}^i\left(\eta_{0}-\lambda\right)\equiv
C(\lambda)\,.
\end{align}
In addition to the initial conditions for $k^0_{(r)}$, $r=1,2$,
given by equation (\ref{a32}), we can impose, first of all,
\begin{equation} \label{a35}
x_{(0)}^i(0)=x_{(1)}^i(0)=x_{(2)}^i(0)=\eta_{(1)}(0)=\eta_{(2)}(0)=0\;,\quad
\eta_{(0)}(0)=\eta_{0}\;;
\end{equation}
the initial conditions for $k_{(1)}^i$ and $k_{(2)}^i$, instead,
must be chosen taking into account the null normalization
condition. Imposing $k^\mu k_\mu=0$ for $\lambda=0$ is indeed a
necessary and sufficient condition to get a null geodesic, once
$k^\mu_{\phantom{\mu};\nu}k^\nu=0$ is solved \cite{geodetiche1}.
Assuming
\begin{equation}\label{a36}
k_{(1)}^i(0)=ye^i
\end{equation}
and imposing $k^\mu(0)k_\mu(0)=0$ to first order, we obtain
\begin{equation}\label{a37}
k_{(0)}^i(0)k_{(0)}^j(0)\left(-2\psi_{(1)}\left(\boldsymbol{x}_{(0)}(\eta_0
\right)\delta_{ij}
+\chi_{(1)ij}\left(\boldsymbol{x}_{(0)},\eta_0\right)\right)+
2k_{(0)}^i(0)k_{(1)}^j(0)\delta_{ij}=2k_{(0)}^0(0)k_{(1)}^0(0)=0\;,
\end{equation}
from which we have
\begin{equation}\label{a38}
y=\psi_{(1)}\left(\boldsymbol{x}_{(0)},\eta_0\right)-
\frac12\chi_{(1)ij}\left(\boldsymbol{x}_{(0)},\eta_0\right)e^ie^j=\frac{5}{3}\varphi\left(\boldsymbol{x}_{(0)}(0)\right)+\frac{\eta_{0}^2}{6}e^ie^j\varphi_{,ij}\left(\boldsymbol{x}_{(0)}(0)\right)\;.
\end{equation}
We are now in a position to find the zeroth, the  first and the
second-order quantities necessary to define the photon trajectory
\begin{gather}\label{a39}
\eta_{(0)}=\eta_{0}-\lambda\;, \qquad x_{(0)}^i=\lambda e^i\;;\\
\label{a41}
k_{(1)}^{0}(\lambda)=\frac{1}{3}\left(\varphi(\boldsymbol{x}_{(0)}(\lambda))-\varphi(\boldsymbol{x}_{(0)}(0))\right)+\frac{1}{3}(\eta_0-\lambda')e^i\varphi_{,i}(\boldsymbol{x}_{(0)}(\lambda'))\bigg\vert^{\lambda'=\lambda}_{\lambda'=0}\;,
\\\label{a43}
k_{(1)}^i\left(\lambda\right)=ye^i-2\int_{0}^{\lambda}\varphi^{,i}d\lambda'
+\frac{10}{3}e^i\varphi\left(\boldsymbol{x}_{(0)}(\lambda')\right)
\bigg\vert^{\lambda'=\lambda}_{\lambda'=0}
+\frac{1}{6}\left(\left(\eta_{0}-\lambda'\right)^2\varphi^{,ij}
\left(\boldsymbol{x}_{(0)}(\lambda')\right)e_j
-2\varphi^{,i}(\boldsymbol{x}_{(0)}(\lambda'))(\eta_0-\lambda')\right)
\bigg\vert^{\lambda'=\lambda}_{\lambda'=0}\;,
\\\label{a44}
\eta_{(1)}\left(\lambda\right)=\frac{2}{3}\int_{0}^{\lambda}\varphi\left(\boldsymbol{x}_{(0)}(\lambda')\right)d\lambda'
-\frac{1}{3}\lambda\varphi\left(\boldsymbol{x}_{(0)}(0)\right)-\frac{1}{3}\eta_{0}\lambda
e^i\varphi_{,i}\left(\boldsymbol{x}_{(0)}(0)\right)
+\frac{1}{3}\left(\left(\eta_{0}-\lambda'\right)\varphi\left(\boldsymbol{x}_{(0)}(\lambda')\right)\right)\bigg\vert^{\lambda'=\lambda}_{\lambda'=0}\;,
\end{gather}
\begin{align}\label{a45}
&x_{(1)}^i(\lambda)=y\lambda
e^i-2\int^{\lambda}_{0}\left(\int^{\lambda'}_{0}\varphi^{,i}\left(\boldsymbol{x}_{(0)}(\lambda'')\right)d\lambda''\right)d\lambda'+\frac{1}{3}\varphi^{,i}(\boldsymbol{x}(0))\eta_0\lambda
+\frac{10}{3}e^i\left(\int_{0}^{\lambda}\varphi\left(\boldsymbol{x}_{(0)}(\lambda')\right)
d\lambda'-\lambda\varphi\left(\boldsymbol{x}_{(0)}(0)\right)\right)\nonumber\\&
+\frac{1}{6}\Bigg(-\eta_{0}^2\lambda\varphi^{,ij}\left(\boldsymbol{x}_{(0)}(0)\right)e_j+\left(\left(\eta_{0}-\lambda'\right)^2\varphi^{,i}\left(\boldsymbol{x}_{(0)}(\lambda')\right)\right)\vert^{\lambda'=\lambda}_{\lambda'=0}
\Bigg)\;;
\end{align}
\begin{gather}\label{a47}
k_{(2)}^0(\lambda)=\int_{0}^{\lambda}C(\lambda')d\lambda'\;,
\\\label{a48}
\eta_{(2)}(\lambda)=\int_{0}^{\lambda}\left(\int_{0}^{\lambda'}C(\lambda'')d\lambda''\right)d\lambda'\;.
\end{gather}
Here and afterwards, if the argument of a function is left
unspecified, we mean it is evaluated at
$(\eta_{(0)}(\lambda),x_{(0)}^i(\lambda))$, given by (\ref{a39}):
for instance
$\varphi_{,ij}e^ie^j=\varphi_{,ij}(\boldsymbol{x}_{(0)}(\lambda))e^ie^j=
\frac{d^2\varphi}{d\lambda^2}(\boldsymbol{x}_{(0)}(\lambda))$.
\subsection{The optical scalars}
Let us  now consider the transport equations for the optical
scalars. To solve them, we obviously need to impose some boundary
conditions. One could wonder if they are independent of the
initial conditions we have imposed on the geodesic crossing  the
observer, Eqs. (\ref{a32}), (\ref{a35}), (\ref{a36}) and
(\ref{a38}) (see Figure \ref{raggio}).
\begin{figure}
\epsfxsize=3in
\begin{center}
\leavevmode \epsffile{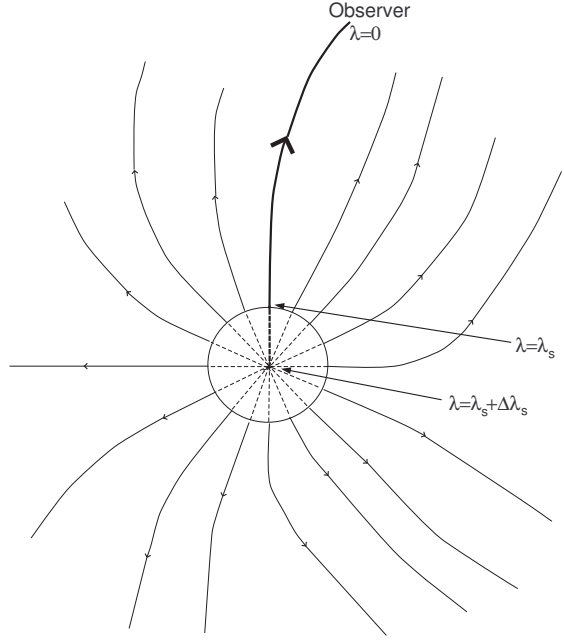}
\end{center}
\caption{A congruence of photons emitted by a spherically
symmetric source: one ray reaches the observer.} \label{raggio}
\end{figure}
Therefore, in order to understand which conditions must be
adopted, let us suppose that our source emits isotropically (more
precisely, suppose that an observer comoving with the source sees
it emitting isotropically). Consider, then, a set
$(\widetilde{\eta},\widetilde{x}^i)$ of locally inertial
coordinates around the source, associated with the tetrad
\eq\left\{e_{(0)},\,e_{(1)},\,e_{(2)},\,e_{(3)}\right\}\,,\quad
e_{(0)}=u^\mu\frac{\partial}{\partial x^\mu}\, ,\feq where
$e_{(0)},\,e_{(1)},\,e_{(2)},\,e_{(3)}$ are tangent to the
space-time in the event corresponding to the center  of the
source; in particular, $u^\mu\frac{\partial}{\partial
x^\mu}=\frac{\partial}{\partial\eta}$ is the 4-velocity of the
center of the source. In other words, using tilded indices for the
components in the coordinates $(\widetilde{\eta},\widetilde{x}^i)$
and indicating with $\widetilde{x}_c$ the coordinates of the
center of the source, we require that,  in a neighborhood of
$\widetilde{x}_c$,
\begin{equation}
g_{\widetilde{\mu}\widetilde{\nu}}(\widetilde{x})=\eta_{\widetilde{\mu}\widetilde{\nu}}+
{\cal O}((\widetilde{x}-\widetilde{x}_c)^2)\quad\mbox{and}\quad
\frac{\partial}{\partial\widetilde{x}^{\mu}}\Big\vert_{\widetilde{x}=\widetilde{x}_c}=
e_{(\mu)}
\end{equation} (this yields, in particular,
$\Gamma^{\widetilde{\mu}}_{\widetilde{\nu}\widetilde{\sigma}}(\widetilde{x})=
{\cal O}(\widetilde{x}-\widetilde{x}_c)$). If we consider the
geodesics on which we imposed (\ref{a32}), (\ref{a35}),
(\ref{a36}) and  (\ref{a38}), in the center of the source the wave
4-vector reads \eq
k^{\widetilde{\mu}}(\lambda_s+\Delta\lambda_s)=\omega(1,m^i)\;,\feq
with \begin{equation}\omega=-u_\mu
k^\mu(\lambda_s+\Delta\lambda_s)=-1+k^0_{(1)}(\lambda_s+\Delta\lambda_s)+
k^0_{(2)}(\lambda_s+\Delta\lambda_s)\;:\end{equation} $m^ie_{(i)}$
(with $m^{i}m^{j}\delta_{ij}=1$) is of course
the direction in which the photon is emitted.\\
If the emission is isotropic, the wave 4-vector of a photon
emitted in a generic direction $n^{i}e_{(i)}$ (with
$n^{i}n^{j}\delta_{ij}=1$) is given by \eq\label{messyeq}
k^{\widetilde{\mu}}(\lambda_s+\Delta\lambda_s,\boldsymbol{n})
=\omega(1,n^{i})\;.\feq As we now have at our disposal the initial
conditions (i.e. at $\lambda=\lambda_s+\Delta\lambda_s$) for all
the geodesics of the congruence, let us reconstruct it in a
neighborhood of the source. First of all, let us note that ${\cal
O}(\lambda-(\lambda_s+\Delta\lambda_s))= {\cal
O}(\widetilde{x}-\widetilde{x}_c)$: by definition, indeed, we have
$\frac{d\widetilde{x}^\mu}{d\lambda}=k^{\widetilde{\mu}}$, which
yields
\begin{equation}
\widetilde{x}^\mu-\widetilde{x}_c^\mu=k^{\widetilde{\mu}}(\lambda_s+\Delta\lambda_s,\boldsymbol{n})\vert_{\boldsymbol{n}=
\frac{\widetilde{\boldsymbol{x}}-\widetilde{\boldsymbol{x}}_c}{\|\widetilde{\boldsymbol{x}}-\widetilde{\boldsymbol{x}}_c\|}}(\lambda-(\lambda_s+\Delta\lambda_s))+
{\cal O}((\lambda-(\lambda_s+\Delta\lambda_s))^2)\;.\end{equation}
Using therefore \eq
\Gamma^{\widetilde{\mu}}_{\widetilde{\nu}\widetilde{\sigma}}(\lambda)={\cal
O}(\lambda-(\lambda_s+\Delta\lambda_s))\;, \feq it is immediate to
get \eq
k^{\widetilde{\mu}}(\lambda,\boldsymbol{n})=k^{\widetilde{\mu}}(\lambda_s+\Delta\lambda_s,\boldsymbol{n})+{\cal
O}(\lambda-(\lambda_s+\Delta\lambda_s))^2\;,\feq which implies
\begin{equation}\label{messyeq2}
\widetilde{x}^\mu-\widetilde{x}_c^\mu=k^{\widetilde{\mu}}(\lambda_s+\Delta\lambda_s,\boldsymbol{n})\vert_{\boldsymbol{n}=
\frac{\widetilde{\boldsymbol{x}}-\widetilde{\boldsymbol{x}}_c}{\|\widetilde{\boldsymbol{x}}-\widetilde{\boldsymbol{x}}_c\|}}(\lambda-(\lambda_s+\Delta\lambda_s))+
{\cal O}((\lambda-(\lambda_s+\Delta\lambda_s))^3)\;;\end{equation}
thus we have \eq
k^{\widetilde{\mu}}(x)=\omega(1,n^i)\vert_{\boldsymbol{n}=
\frac{\widetilde{\boldsymbol{x}}-\widetilde{\boldsymbol{x}}_c}{\|\widetilde{\boldsymbol{x}}-\widetilde{\boldsymbol{x}}_c\|}}
+{\cal O}((\widetilde{x}-\widetilde{x}_c)^2)\;.\feq From this
equation we easily obtain
\begin{gather} \theta(\widetilde{x})=\nabla_{\widetilde{\mu}}
k^{\widetilde{\mu}}=\frac{2\omega}{\|\widetilde{\boldsymbol{x}}-\widetilde{\boldsymbol{x}}_c\|}+{\cal O}(\widetilde{x}-\widetilde{x}_c)\;,\\
\sigma(\widetilde{x})={\cal O}(\widetilde{x}-\widetilde{x}_c)\;,
\end{gather}
which, along a geodesic, become
\begin{gather}
\theta(\lambda)=\frac{2}{\lambda-(\lambda_s+\Delta\lambda_s)}+{\cal O}(\lambda-(\lambda_s+\Delta\lambda_s))\;,\\
\sigma(\lambda)={\cal
O}(\lambda-(\lambda_s+\Delta\lambda_s))\end{gather} (use
(\ref{messyeq}) and (\ref{messyeq2})). It is therefore clear that
the boundary conditions to impose on the transport equations for
the optical scalars are
\begin{gather}
\theta_{(0)}(\lambda)\sim\frac{2}{\lambda-(\lambda_s+\Delta\lambda_s)}
\qquad\mbox{as}\quad\lambda\to\lambda_s+\Delta\lambda_s\;,\label{theta0}\\
\sigma_{(0)}(\lambda_s+\Delta\lambda_s)=0\;,\label{sigma0}\\\label{altrecondcont}
\sigma_{(1)}(\lambda_s+\Delta\lambda_s)=\theta_{(1)}(\lambda_s+\Delta\lambda_s)=\theta_{(2)}(\lambda_s+\Delta\lambda_s)=0\;.
\end{gather}
In particular, from (\ref{sigma0}) we immediately get (see
equation (\ref{a18}))
\begin{equation}\label{a49}
\sigma_{(0)}(\lambda)=0\;,\quad\forall\lambda\;;
\end{equation}
thus equation (\ref{a17}) becomes, to zeroth order,
\begin{equation}\label{a50}
\frac{d\theta_{(0)}}{d\lambda}=-\frac{\theta_{(0)}^2}{2}\;,
\end{equation}
which yields, once Eq. (\ref{theta0}) is imposed,
\begin{equation}\label{a51}
\theta_{(0)}\left(\lambda,\Delta\lambda_s,\lambda_s\right)
=\frac{2}{\lambda-\lambda_s-\Delta\lambda_s}\;.
\end{equation}
Similarly, perturbing (\ref{a17}) to first and second order, we
obtain
\begin{equation}\label{a52}
\frac{d\theta_{(1)}}{d\lambda} =
-\theta_{(0)}\theta_{(1)}-R_{(1)\mu\nu}k_{(0)}^{\mu}k_{(0)}^{\nu}\;,
\end{equation}
from which we have, using Eq.  (\ref{altrecondcont}),
\begin{equation}\label{a53}
\theta_{(1)}\left(\lambda,\Delta\lambda_s,\lambda_s\right)=
\frac{1}{\left(\lambda-\lambda_s-\Delta\lambda_s\right)^2}\int_{\lambda}^{\lambda_s+\Delta\lambda_s}\left(\lambda'-\lambda_s-\Delta\lambda_s\right)^2\left(R_{(1)\mu\nu}k_{(0)}^{\mu}k_{(0)}^{\nu}\right)(\lambda')d\lambda'\;,
\end{equation}
and
\begin{multline}\label{a54}
\frac{d\theta_{(2)}}{d\lambda}=-\theta_{(0)}\theta_{(2)}-B\left(\lambda,\Delta\lambda_s,\lambda_s\right)\;,\\
B\left(\lambda,\Delta\lambda_s,\lambda_s\right)=R_{(2)\mu\nu}k_{(0)}^{\mu}k_{(0)}^{\nu}+2R_{(1)\mu\nu}k_{(0)}^{\mu}k_{(1)}^{\nu}+
R_{(1)\mu\nu,\alpha}x_{(1)}^{\alpha}k_{(0)}^{\mu}k_{(0)}^{\nu}
+\frac{\theta_{(1)}^2}{2}+2\vert\sigma_{(1)}\vert^2\;,
\end{multline}
with

\eq R_{(1)00}=3{{\psi_{(1)}}}^{\prime\prime}\, ,\feq \eq
R_{0i(1)}={\frac12}\left({\chi_{(1)}}_{i,k}^{k}\right)^\prime
+2\,\left({\psi_{(1)}}_{,i}\right)^\prime\, , \feq
\begin{equation}
R_{(1)ij}=\delta_{ij}\nabla^2{\psi_{(1)}}
-{\frac12}\nabla^2{{\chi_{(1)}}_{ij}}
+{\frac12}{{\chi_{(1)}}^{k}_{i,kj}}
+{\frac12}{{\chi_{(1)}}^{k}_{j,ki}}
-\delta_{ij}{{\psi_{(1)}}}^{\prime\prime} +{{\psi_{(1)}}}_{,ij}
+{\frac12}{{\chi_{(1)}}_{ij}}^{\prime\prime}\, ,
\end{equation}
\begin{equation}\label{a60}
\theta_{(2)}\left(\lambda,\lambda_s\right)=
\frac{1}{\left(\lambda-\lambda_s\right)^2}\int_{\lambda}^{\lambda_s}\left(\lambda'-\lambda_s\right)^2B\left(\lambda',\lambda_s\right)d\lambda'\;,
\end{equation}
\begin{equation}
B\left(\lambda,\lambda_s\right)=R_{(2)\mu\nu}k_{(0)}^{\mu}k_{(0)}^{\nu}+2R_{(1)\mu\nu}k_{(0)}^{\mu}k_{(1)}^{\nu}+
R_{(1)\mu\nu,\alpha}x_{(1)}^{\alpha}k_{(0)}^{\mu}k_{(0)}^{\nu}
+\frac{\theta_{(1)}^2}{2}+2\vert\sigma_{(1)}\vert^2\;,
\end{equation}

\begin{equation}
R_{(2)00}={\frac14}\left({\chi_{(1)}}_{kl}\right)^{\prime}
\left({{\chi_{(1)}}^{kl}}\right)^{\prime}
+{\frac12}{\chi_{(1)}}^{kl}\left({\chi_{(1)}}_{kl}\right)^{\prime\prime}
+3\left({{{\psi_{(1)}}}^{\prime}}\right)^{2}
+6{\psi_{(1)}}{{\psi_{(1)}}}^{\prime\prime}
+{\frac32}{{\psi_{(2)}}}^{\prime\prime}\, ,
\end{equation}

\begin{align}& R_{0i(2)}=
{\chi_{(1)}}^k_{i}\left({\psi_{(1)}}_{,k}\right)^\prime
-{\frac12}{{\psi_{(1)}}_{,k}}\left({{\chi_{(1)}}^k_{i}}\right)^{\prime}
+{\psi_{(1)}}\left({\chi_{(1)}}^k_{i,k}\right)^\prime
-{\frac12}{\chi_{(1)}}^k_j\left({{\chi_{(1)}}_{i,k}^{j}}\right)^\prime
\nonumber\\&+{\frac14}\left({\chi_{(1)}}_{kj}\right)^{\prime}{{\chi_{(1)}}^{kj}_{\phantom{kj},i}}
-{\frac12}{{\chi_{(1)}}^{kj}}_{,j}\left({{\chi_{(1)}}_{ik}}\right)^{\prime}
+{\frac12}{\chi_{(1)}}_{kj}\left({{\chi_{(1)}}^{kj}}_{,i}\right)^\prime
+{\frac14}\left({{\chi_{(2)}}^k_{i,k}}\right)^\prime
\nonumber\\&+{{\chi_{(1)}}_{i,k}^{k}}{{\psi_{(1)}}}^{\prime}
+4{{\psi_{(1)}}}^{\prime}{\psi_{(1)}}_{,i}
+4{\psi_{(1)}}\left({{\psi_{(1)}}_{,i}}\right)^\prime
+\left({{\psi_{(2)}}_{,i}}\right)^\prime\, ,
\end{align}

\begin{align}
& R_{(2)ij}=\delta_{ij}\left(\nabla{\psi_{(1)}}\right)^2
+{\chi_{(1)}}^{k}_j{{\psi_{(1)}}}_{,ki} +{\chi_{(1)}}^{k}_{i
}{{\psi_{(1)}}}_{,kj}
+2\,\delta_{ij}{\psi_{(1)}}\nabla^2{\psi_{(1)}}
\nonumber\\
&-\delta_{ij}\,{\chi_{(1)}}^{kl}{{\psi_{(1)}}}_{,kl}
+{\frac12}\,\delta_{ij}\nabla^2{\psi_{(2)}}
-{\frac32}{{\psi_{(1)}}}^{,k}{\chi_{(1)}}_{ij,k}
-{\psi_{(1)}}\nabla^2{\chi_{(1)}}_{ij}
\nonumber\\
& +{\frac12}{\chi_{(1)}}^{kl}{\chi_{(1)}}_{ij,kl}
+{\frac12}{{\psi_{(1)}}}_{,k}{{\chi_{(1)}}^{k}_{ i ,j}}
+{\psi_{(1)}}{{\chi_{(1)}}_{i,kj}^k}
-{\frac12}{\chi_{(1)}}^{k}_l{{\chi_{(1)}}_{ik,lj}}
\nonumber\\
&-{\frac12}\left({\chi_{(1)}}^{k}_{i}\right)^{\prime}\left({\chi_{(1)}}_{kj
}\right)^{\prime}
+{\frac12}{{\psi_{(1)}}}_{,k}{{\chi_{(1)}}^{k}_{j,i }}
+{\frac12}{{\chi_{(1)}}_{ik,l}}{{\chi_{(1)}}^{k,l}_{j }}
+{\psi_{(1)}}{{\chi_{(1)}}^{k}_{j,ki }}
\nonumber\\
&-{\frac12}{{\chi_{(1)}}^k_{i,l}}{{\chi_{(1)}}^{l}_{j,k }}
-{\frac12}{\chi_{(1)}}^{kl}{{\chi_{(1)}}_{jl,ki}}+{\frac14}{{\chi_{(1)}}^{kl}}_{,i}{{\chi_{(1)}}_{kl,j}}
-\delta_{ij}{{\psi_{(1)}}}_{,k}{{\chi_{(1)}}^{kl}_{  }}_{,l}\nonumber\\
& +{\frac12}{{\chi_{(1)}}^{kl}_{
}}_{,l}\left({{\chi_{(1)}}_{ij,k}} -{{\chi_{(1)}}_{ik,j}}
-{{\chi_{(1)}}_{jk,i}}\right) +{\frac12}{\chi_{(1)}}^{kl}_{
}{{\chi_{(1)}}_{kl,ij}} -{\frac14}\nabla^2{{\chi_{(2)}}_{ij}}\nonumber\\
& +{\frac14}{{\chi_{(2)}}^{k}_{i,kj }}
+{\frac14}{{\chi_{(2)}}^{k}_{j,ki }}
+\delta_{ij}\left({{\psi_{(1)}}}^{\prime}\right)^{2}
+{{\chi_{(1)}}^{k}_{j,k}}{{\psi_{(1)}}}_{,i} +{{\chi_{(1)}}^{k}_{
i,k }}{{\psi_{(1)}}}_{,j} \nonumber\\
&+3{{\psi_{(1)}}}_{,i}{{\psi_{(1)}}}_{,j}
+2{\psi_{(1)}}{{\psi_{(1)}}}_{,ij}
-{\frac12}\delta_{ij}{{\psi_{(2)}}}^{\prime\prime}
+{\frac12}{{\psi_{(2)}}}_{,ij}
+{\frac12}{{\psi_{(1)}}}^{\prime}\left({\chi_{(1)}}_{ij}\right)^{\prime}
+{\frac14}\left({\chi_{(2)}}_{ij}\right)^{\prime\prime}\, ,
\end{align}

or

\begin{gather}\label{a55}R_{(1)00}(\lambda)=\frac{1}
{3}\nabla^2{\varphi}\;,\\ \label{a56}
R_{(2)00}(\lambda)=-\frac{5}{9}\nabla^2\left(\varphi^2\right)+\frac{5}{6}
{\varphi}^{,a}{\varphi}_{,a} +\frac{20}{9}
\varphi\nabla^2{\varphi}+(\eta_0-\lambda)^2\left( -\frac{1}{63}
{\varphi}^{,ab}{\varphi}_{,ab} +\frac{1}{14}
\left(\nabla^2{\varphi}\right)^2\right)\;,\\
R_{(1)0i}(\lambda)=R_{(2)0i}(\lambda)=0\;,\label{a57}\\
\label{a58} R_{(1)ij}(\lambda)=\frac{5}{3}
\delta_{ij}\nabla^2{\varphi} +\frac{4}{3}{\varphi}_{,ij}\;,\end{gather}
\begin{align}\label{a59}
&R_{(2)ij}(\lambda)=\delta_{ij}\Bigg[\frac{55}{18}
{\varphi}^{,a}{\varphi}_{,a}+\frac{50}{9}
\nabla^2{\varphi}\varphi\delta_{ij}+(\eta_0-\lambda)^{2}\left(\frac{13}{63}
{\varphi}^{,ab}{\varphi}_{,ab}+\frac{1}{14}
\left(\nabla^2{\varphi}\right)^2\right)\nonumber\\&+\frac{1}{84}
(\eta_0-\lambda)^{4}\left({\varphi}^{,abc}{\varphi}_{,abc}
-{\varphi}^{,ab}_{\phantom{ab},b}{\varphi}^{,c}_{\phantom{a},ca}
+{\varphi}_{,ab}{\varphi}^{,abc}_{\phantom{abc},c}
-\delta_{ij}\nabla^2{\varphi}{\varphi}^{,bc}_{\phantom{ab},bc}
\right)-\frac{25}{9}\nabla^2\left(\varphi^2\right)\Bigg]\nonumber\\&
+\frac{1}{4}\left(\partial_\eta^2-\nabla^2\right)\pi_{ij}
+\,\frac{20}{3}{\varphi}_{,i}{\varphi}_{,j} +\,\frac{40}{9}
\varphi{\varphi}_{,ij}-\frac{20}{9}\partial_i\partial_j\left(\varphi^2\right)
% +\,\frac{2}{3}\tau_{ij}\nonumber\\&
+(\eta_0-\lambda)^{2}\left(-\frac{38}{63}\
{\varphi}^{,a}_{\phantom{a},i}{\varphi}_{,aj}%-\frac{1}{3}
%\tau^{ij,a}_{\phantom{abc},a}
-\frac{11}{63}\nabla^2{\varphi}{\varphi}_{,ij} %+\frac{1}{7}
%S_{ij}
\right)\nonumber\\& +\frac{1}{84}(\eta_0-\lambda)^{4}
\Bigg[-{\varphi}^{,a}_{\phantom{a},ai}{\varphi}^{,b}_{\phantom{a},bj}
+{\varphi}^{,ab}{\varphi}_{,abij}
+\nabla^2{\varphi}{\varphi}^{,b}_{\phantom{a},bij}
-3{\varphi}^{,ab}_{\phantom{ab},i}{\varphi}_{,abj}\nonumber\\&+2\bigg(-{\varphi}_{,aj}{\varphi}^{,ab}_{\phantom{ab},bi}
-{\varphi}_{,ai}{\varphi}^{,ab}_{\phantom{ab},bj}
+{\varphi}^{,ab}_{\phantom{ab},ab}{\varphi}_{,ij}\bigg)+4{\varphi}^{,ab}_{\phantom{ab},b}{\varphi}_{,aij}\Bigg] \;.
\end{align}
Equation (\ref{a54}), using Eq. (\ref{altrecondcont}), is
therefore solved by
\begin{equation}\label{a60}
\theta_{(2)}\left(\lambda,\Delta\lambda_s,\lambda_s\right)=
\frac{1}{\left(\lambda-\lambda_s-\Delta\lambda_s\right)^2}\int_{\lambda}^{\lambda_s+\Delta\lambda_s}\left(\lambda'-\lambda_s-\Delta\lambda_s\right)^2B\left(\lambda',\Delta\lambda_s,\lambda_s\right)d\lambda'\;.
\end{equation}
As $B\left(\lambda',\Delta\lambda_s,\lambda_s\right)$ depends on
$\sigma_{(1)}$ too, we must solve the first of Eq. (\ref{a18}),
perturbed to first order,
\begin{equation}\label{a61}
\frac{d\sigma_{(1)}}{d\lambda}=-\theta_{(0)}\sigma_{(1)}+
R_{(1)\sigma\mu\nu\delta}k_{(0)}^{\sigma}k_{(0)}^{\delta}\bar{t}_{(0)}^{\mu}\bar{t}_{(0)}^{\nu}\;,
\end{equation}
\begin{equation}
R_{(1)i0j0}=\delta_{ij}{\psi^{\prime\prime}_{(1)}}
-{\frac12}\left({\chi_{(1)}}_{ij}\right)^{\prime\prime}\, ,
\end{equation}
\begin{equation}
R_{(1)0ijk}=\delta_{ik}\left({\psi_{(1)}}_{,j}\right)^{\prime}
-\delta_{ij}\left({\psi_{(1)}}_{,k}\right)^{\prime}
+{\frac12}\left({\chi_{(1)}}_{ij,k}\right)^\prime
-{\frac12}\left({\chi_{(1)}}_{ik,j}\right)^\prime\, ,
\end{equation}
with
\begin{equation}
R_{(1)i0j0}=\delta_{ij}{\psi^{\prime\prime}_{(1)}}
-{\frac12}\left({\chi_{(1)}}_{ij}\right)^{\prime\prime}\, ,
\end{equation}
\begin{equation}
R_{(1)0ijk}=\delta_{ik}\left({\psi_{(1)}}_{,j}\right)^{\prime}
-\delta_{ij}\left({\psi_{(1)}}_{,k}\right)^{\prime}
+{\frac12}\left({\chi_{(1)}}_{ij,k}\right)^\prime
-{\frac12}\left({\chi_{(1)}}_{ik,j}\right)^\prime\, ,
\end{equation}
\begin{equation}
R_{(1)ijkl}=\delta_{jl}{\psi_{(1)}}_{,ik}
-\delta_{jk}{\psi_{(1)}}_{,il} -\delta_{il}{\psi_{(1)}}_{,jk}
+\delta_{ik}{\psi_{(1)}}_{,jl} -{\frac12}{\chi_{(1)}}_{ik,jl}
+{\frac12}{\chi_{(1)}}_{il,jk} +{\frac12}{\chi_{(1)}}_{jk,il}
-{\frac12} {\chi_{(1)}}_{jl,ik}\, ,
\end{equation}
or
\begin{multline}
R_{(1)i0j0}=\frac{1}{3}\varphi_{,ij}\;,\quad R_{(1)0ijk}=0\;,
R_{(1)ijkl}=\frac{5}{3}\left(\varphi_{,ik}\delta_{jl}+\varphi_{,jl}\delta_{ik}-\varphi_{,jk}\delta_{il}-\varphi_{,il}\delta_{jk}\right)\;,
\end{multline}
with $\bar{t}_{(0)}^{\mu}$ obeying, to zeroth order, the second of
Eq. (\ref{a18}):
\begin{multline}\label{a62}
\bar{t}_{(0)}^{\mu}=\frac{v^{\mu}+iw^{\mu}}{\sqrt{2}},\quad
v^{\mu}=\left(0,v^i\right),\quad \delta_{ij}v^iv^j=1,\quad
w^{\mu}=\left(0,w^i\right),\quad \delta_{ij}w^iw^j=1,\\
\delta_{ij}v^ie^j=\delta_{ij}w^ie^j=\delta_{ij}w^iv^j=0,\quad
\frac{dv^i}{d\lambda}=\frac{dw^i}{d\lambda}=0\;.
\end{multline}
Imposing equation (\ref{altrecondcont}), the solution reads
\begin{equation}\label{a63}
\sigma_{(1)}\left(\lambda,\Delta\lambda_s,\lambda_s\right)=
-\frac{1}{\left(\lambda-\lambda_s-\Delta\lambda_s\right)^2}\int_{\lambda}^{\lambda_s+\Delta\lambda_s}\left(\lambda'-\lambda_s-\Delta\lambda_s\right)^2\left(R_{(1)\sigma\mu\nu\delta}k_{(0)}^{\sigma}k_{(0)}^{\delta}\bar{t}_{(0)}^{\mu}\bar{t}_{(0)}^{\nu}\right)\left(\lambda'\right)d\lambda'\;.
\end{equation}
\subsection{The luminosity distance} To get the relation
 between the physical radius of
the source $R$ and $\Delta\lambda_s$ let us use again a set
$\left(\widetilde{\eta},\widetilde{x}^i\right)$ of locally
inertial coordinates in a neighborhood of the center of the source
and let us define $\hat{k}^{\mu}=a^{-2}k^{\mu}$ (this corresponds
to Eq. (\ref{a6}) in the old affine parametrization). We therefore
have \eq \label{a64}
\left(\hat{g}_{\mu\nu}\hat{u}^{\mu}\hat{k}^{\nu}\right)\left(\lambda_s\right)
=\frac{1}{a\left(\eta\left(\lambda_s\right)\right)}\left(1-k_{(1)}^0\left(\lambda_s\right)
-k_{(2)}^0\left(\lambda_s\right)\right)=
-\frac{1}{a\left(\eta\left(\lambda_s\right)\right)^2}\left(\frac{d\widetilde{\eta}}{d\lambda}\left(\lambda_s\right)\right)
\feq (in particular, we have used Eq. (\ref{a1}) and the new
definition of $\hat{k}^{\mu}$). For the second of Eq. (\ref{a7})
yields
$d\widetilde{\eta}^2=\delta_{ij}d\widetilde{x}^id\widetilde{x}^j$,
our relation reads
\begin{equation} \label{a65}
R=\sqrt{\delta_{ij}d\widetilde{x}^id\widetilde{x}^j}=\vert\Delta\widetilde{\eta}\vert=a\left(\eta\left(\lambda_s\right)\right)\Delta\lambda_s\left(1-k_{(1)}^0\left(\lambda_s\right)
-k_{(2)}^0\left(\lambda_s\right)\right)\;.
\end{equation}
We can now solve for the amplitude $A$
\begin{equation}\label{a66}
\left(Aa\right)\left(\lambda_s\right)=\left(Aa\right)\left(0\right)e^{-\frac{1}{2}\int_{0}^{\lambda_s}\theta\left(\lambda\right)d\lambda}\;;
\end{equation}
and  find
\begin{align}\label{a67}
d_L=&\exp\left[-\frac{1}{2}\int_{0}^{\lambda_s}
\left(\theta_{(0)}\left(\lambda,\Delta\lambda_s,\lambda_s\right)+
\theta_{(1)}\left(\lambda,\Delta\lambda_s,\lambda_s\right)+
\theta_{(2)}\left(\lambda,\Delta\lambda_s,\lambda_s\right)\right)
d\lambda\right]\nonumber\\
&\times\left(1+\widetilde{z}\left(\lambda_s\right)\right)a_0\Delta\lambda_s
\left(1-k_{(1)}^0\left(\lambda_s\right)-k_{(2)}^0\left(\lambda_s\right)
\right)\;,
\end{align}
where we have used $a(\eta(0))=a(\eta_0)=a_0$. Taking the limit
for $\Delta\lambda_s\to0$, noting that
\begin{equation}\label{a68}
e^{-\frac{1}{2}\int_{0}^{\lambda_s}\left(\theta_{(0)}\left(\lambda,\Delta\lambda_s,\lambda_s\right)\right)d\lambda}=\frac{\lambda_s+\Delta\lambda_s}{\Delta\lambda_s}\;,
\end{equation}
and expanding the exponential, we get
\begin{multline}\label{a69}
d_L=\left(1+\widetilde{z}\left(\lambda_s\right)\right)a_0\lambda_s
\Bigg[1-\frac{1}{2}\int_{0}^{\lambda_s}\theta_{(1)}\left(\lambda,\lambda_s\right)d\lambda+\frac{1}{8}\left(\int_{0}^{\lambda_s}\theta_{(1)}\left(\lambda,\lambda_s\right)d\lambda\right)^2\\
-\frac{1}{2}\int_{0}^{\lambda_s}\theta_{(2)}\left(\lambda,\lambda_s\right)d\lambda
-k_{(1)}^0\left(\lambda_s\right)-k_{(2)}^0\left(\lambda_s\right)+\frac{1}{2}k_{(1)}^0\left(\lambda_s\right)\int_{0}^{\lambda_s}\theta_{(1)}\left(\lambda,\lambda_s\right)d\lambda\Bigg]\;,
\end{multline}
where $\theta_{(1)}\left(\lambda,\lambda_s\right)$ and
$\theta_{(2)}\left(\lambda,\lambda_s\right)$ stand for
$\theta_{(1)}\left(\lambda,\Delta\lambda_s=0,\lambda_s\right)$ and
$\theta_{(2)}\left(\lambda,\Delta\lambda_s=0,\lambda_s\right)$. In
order to express the right-hand side of equation (\ref{a69}) as a
function of the redshift we can invert, order by order, the
following equation (to get it, we have used Eq. (\ref{a23}),
(\ref{a32}) and (\ref{a35})):
\begin{align}\label{a70}
&1+\widetilde{z}\left(\lambda_s\right)=\frac{a_0}{a\left(\eta\left(\lambda_s\right)\right)}
\frac{\left(u_{\mu}k^{\mu}\right)\left(\lambda_s\right)}{\left(u_{\mu}k^{\mu}\right)\left(0\right)}=\nonumber\\&
\frac{a_0\left(1-\left(k_{(1)}^0\left(\lambda_s\right)+k_{(2)}^0\left(\lambda_s\right)\right)\right)}
{a\left(\eta_{(0)}\left(\lambda_s\right)\right)
+a'\left(\eta_{(0)}\left(\lambda_s\right)\right)\left(\eta_{(1)}\left(\lambda_s\right)+\eta_{(2)}\left(\lambda_s\right)\right)
+\frac{1}{2}a''\left(\eta_{(0)}\left(\lambda_s\right)\right)\eta_{(1)}\left(\lambda_s\right)^2}=\nonumber\\&
\frac{a_0}{a\left(\eta_{(0)}\left(\lambda_s\right)\right)}\left(1+T_1(\lambda_s)+T_2(\lambda_s)\right)\;,\end{align}
\begin{equation}\label{t1expr}
T_1\left(\lambda\right)=\left(-\frac{a'(\eta_{(0)})}{a(\eta_{(0)})}\eta_{(1)}-k_{(1)}^0\right)\left(\lambda\right)\;,
\feq \eq
T_2\left(\lambda\right)=\Bigg(-\frac{a'(\eta_{(0)})}{a(\eta_{(0)})}\eta_{(2)}-k_{(2)}^0-\frac{a''(\eta_{(0)})}{2a(\eta_{(0)})}\eta_{(1)}^2+\left(\frac{a'(\eta_{(0)})}{a(\eta_{(0)})}\eta_{(1)}\right)^2+\frac{a'(\eta_{(0)})}{a(\eta_{(0)})}\eta_{(1)}k_{(1)}^0
\Bigg)(\lambda)\label{t2expr}\;.\end{equation} Adopting thus the
definition
\begin{equation}\label{a71}
\lambda_s=\lambda_{(0)}+\lambda_{(1)}+\lambda_{(2)}
\end{equation}
and inserting it in the right-hand side of equation (\ref{a70}),
we obtain \eq 1+\widetilde{z}=
\frac{a_0\left(1+T_1\left(\lambda_{(0)}\right)+T_2\left(\lambda_{(0)}\right)+T_1'\left(\lambda_{(0)}\right)\lambda_{(1)}\right)}
{a\left(\eta_0-\lambda_{(0)}\right)-a'\left(\eta_0-\lambda_{(0)}\right)\left(\lambda_{(1)}+\lambda_{(2)}\right)+a''\left(\eta_0-\lambda_{(0)}\right)\frac{\lambda_{(1)}^2}{2}}=\nonumber\feq
\begin{align}\label{a72}&=\frac{a_0}{a\left(\eta_0-\lambda_{(0)}\right)}\Bigg[1+\frac{a'\left(\eta_0-\lambda_{(0)}\right)\left(\lambda_{(1)}+\lambda_{(2)}\right)}{a\left(\eta_0-\lambda_{(0)}\right)}-\frac{a''\left(\eta_0-\lambda_{(0)}\right)\lambda_{(1)}^2}{2\,a\left(\eta_0-\lambda_{(0)}\right)}
+T_1\left(\lambda_{(0)}\right)\nonumber\\&+T_2\left(\lambda_{(0)}\right)+T_1'\left(\lambda_{(0)}\right)\lambda_{(1)}+\left(\frac{a'\left(\eta_0-\lambda_{(0)}\right)}{a\left(\eta_0-\lambda_{(0)}\right)}\lambda_{(1)}\right)^2
+\frac{a'\left(\eta_0-\lambda_{(0)}\right)}{a\left(\eta_0-\lambda_{(0)}\right)}\lambda_{(1)}T_1\left(\lambda_{(0)}\right)\Bigg]\;.
\end{align}
Setting then the zeroth order term equal to $1+\tilde{z}$ and
those of higher order to $0$ \cite{sasakidist}, we have
\eq\label{a73}
1+\tilde{z}=\frac{a_0}{a\left(\eta_0-\lambda_{(0)}\right)}\;,\feq
\eq \label{a74}
\lambda_{(1)}(\lambda_{(0)})=-\frac{a\left(\eta_0-\lambda_{(0)}\right)}{a'\left(\eta_0-\lambda_{(0)}\right)}T_1\left(\lambda_{(0)}\right)=
\eta_{(1)}\left(\lambda_{(0)}\right)+\frac{k_{(1)}^0\left(\lambda_{(0)}\right)}{\mathcal{H}\left(\eta_0-\lambda_{(0)}\right)},\quad
\mathcal{H}(\eta)=\frac{a'(\eta)}{a(\eta)}\;,\feq
\begin{align}
\label{a75}
&\lambda_{(2)}(\lambda_{(0)})=-\frac{1}{\mathcal{H}\left(\eta_0-\lambda_{(0)}\right)}\Bigg(T_2\left(\lambda_{(0)}\right)+T_1'\left(\lambda_{(0)}\right)\lambda_{(1)}+T_1\left(\lambda_{(0)}\right)\lambda_{(1)}\mathcal{H}\left(\eta_0-\lambda_{(0)}\right)\nonumber\\
&+\left(\mathcal{H}\left(\eta_0-\lambda_{(0)}\right)\lambda_{(1)}\right)^2
-\frac{\lambda_{(1)}^2}{2}\frac{a''\left(\eta_0-\lambda_{(0)}\right)}{a\left(\eta_0-\lambda_{(0)}\right)}\Bigg)\;.
\end{align}
This procedure amounts to identifying the physical
redshift parameter with the redshift in the unperturbed Universe
evaluated at the zeroth affine distance \cite{sasakidist}

\begin{equation}
\widetilde{z}(\lambda_s)=z(\lambda_{(0)})\, .
\end{equation}

If we replace $\lambda_s$ in equation (\ref{a69}) by means of
(\ref{a71}) and we expand in series, we obtain
\begin{align}
&d_L=\frac{a_0^2}{a\left(\eta_0-\lambda_{(0)}\right)}\left(\lambda_{(0)}+\lambda_{(1)}+\lambda_{(2)}\right)
\Bigg[1-\frac{1}{2}\int_{0}^{\lambda_{(0)}}\theta_{(1)}\left(\lambda,\lambda_{(0)}\right)d\lambda
-\frac{1}{2}\lambda_{(1)}\int_{0}^{\lambda_{(0)}}\frac{\partial\theta_{(1)}}{\partial\lambda_s}\left(\lambda,\lambda_s\right)\vert_{\lambda_s=\lambda_{(0)}}d\lambda
\nonumber\\&+\frac{1}{8}\left(\int_{0}^{\lambda_{(0)}}\theta_{(1)}\left(\lambda,\lambda_{(0)}\right)d\lambda\right)^2
-\frac{1}{2}\int_{0}^{\lambda_{(0)}}\theta_{(2)}\left(\lambda,\lambda_{(0)}\right)d\lambda
-k_{(1)}^0\left(\lambda_{(0)}\right)-k_{(2)}^0\left(\lambda_{(0)}\right)-\frac{dk^0_{(1)}}{d\lambda}\left(\lambda_{(0)}\right)\lambda_{(1)}\nonumber\\&+\frac{1}{2}k_{(1)}^0\left(\lambda_{(0)}\right)\int_{0}^{\lambda_{(0)}}\theta_{(1)}\left(\lambda,\lambda_{(0)}\right)d\lambda\Bigg]
=\nonumber\\
&d_{L(0)}\Bigg\{1+\left[\frac{\lambda_{(1)}}{\lambda_{(0)}}
-\frac{1}{2}\int_{0}^{\lambda_{(0)}}\theta_{(1)}\left(\lambda,\lambda_{(0)}\right)d\lambda
-k_{(1)}^0\left(\lambda_{(0)}\right)\right]\nonumber\\&
+\bigg[\frac{\lambda_{(2)}}{\lambda_{(0)}}
-\frac{1}{2}\lambda_{(1)}\int_{0}^{\lambda_{(0)}}\frac{\partial\theta_{(1)}}{\partial\lambda_s}\left(\lambda,\lambda_s\right)\vert_{\lambda_s=\lambda_{(0)}}d\lambda
-\frac{dk^0_{(1)}}{d\lambda}\left(\lambda_{(0)}\right)\lambda_{(1)}-\frac{\lambda_{(1)}}{\lambda_{(0)}}k_{(1)}^0\left(\lambda_{(0)}\right)
+\frac{1}{8}\left(\int_{0}^{\lambda_{(0)}}\theta_{(1)}\left(\lambda,\lambda_{(0)}\right)d\lambda\right)^2\nonumber\\&
-\frac{1}{2}\int_{0}^{\lambda_{(0)}}\theta_{(2)}\left(\lambda,\lambda_{(0)}\right)d\lambda
-k_{(2)}^0\left(\lambda_{(0)}\right)
+\frac{1}{2}k_{(1)}^0\left(\lambda_{(0)}\right)\int_{0}^{\lambda_{(0)}}\theta_{(1)}\left(\lambda,\lambda_{(0)}\right)d\lambda-\frac{1}{2}\frac{\lambda_{(1)}}{\lambda_{(0)}}\int_{0}^{\lambda_{(0)}}\theta_{(1)}\left(\lambda,\lambda_{(0)}\right)d\lambda
\bigg]\,\Bigg\}\;. \label{a76}\end{align} The zeroth order
luminosity distance is given by
\begin{equation}\label{a77bis}
d_{L(0)}=\frac{a_0^2}{a\left(\eta_0-\lambda_{(0)}\right)}\lambda_{(0)}=\frac{2}{{\cal
H}_0}\left(1+\widetilde{z}-\sqrt{1+\widetilde{z}}\right)\;,
\end{equation}
where we have used
$a\left(\eta\right)=a_0\left(\frac{\eta}{\eta_0}\right)^2$ and the
inverse of Eq. (\ref{a73}),
\begin{equation}\label{a78}
\frac{\lambda_{(0)}}{\eta_0}=1-\frac{1}{\sqrt{1+\widetilde{z}}}\;.
\end{equation}
Besides, it's worth noting that, in order to get equation
(\ref{a76}), we have used
$\theta_{(1)}(\lambda_{(0)},\lambda_{(0)})=0$. Making indeed the
assumption that $R_{(1)\mu\nu}k_{(0)}^{\mu}k_{(0)}^{\nu}$,
$B\left(\lambda',\Delta\lambda_s,\lambda_s\right)$ and
$R_{(1)\sigma\mu\nu\delta}k_{(0)}^{\sigma}k_{(0)}^{\delta}\bar{t}_{(0)}^{\mu}\bar{t}_{(0)}^{\nu}$
can be written as a power-series
\eq\sum_{n=0}^{+\infty}c_n(\lambda-\lambda_s)^n\;,\feq
$\theta_{(1)}$, $\theta_{(2)}$ and $\sigma_{(1)}$ take the form
\eq\label{serie}\sum_{n=0}^{+\infty}\frac{c_n(\lambda-\lambda_s)^{n+1}}{n+3}\;.\feq
\section{The deceleration parameter}
\noindent As already anticipated, to get $\widetilde{q}_0$ we can
replace $\lambda_{(0)}/\eta_0=1-1/\sqrt{1+\widetilde{z}}$ in Eq.
(\ref{a76}) and expand in series around $\widetilde{z}=0$ (for
simplicity, we will set  $\eta_0=1$). Defining
\mbox{${{\theta}_{(1)}}^{(0,1)}(0,0)=\frac{\partial\theta}{\partial\lambda_s}(\lambda,\lambda_s)\vert_{\lambda=\lambda_s=0}$}
and
${{\theta}_{(1)}}^{(1,0)}(0,0)=\frac{\partial\theta}{\partial\lambda}(\lambda,\lambda_s)\vert_{\lambda=\lambda_s=0}$
and indicating with a prime the differentiation of $\eta_{(1)}$
and $\eta_{(2)}$  with respect to $\lambda$ (for example
\mbox{$\eta_{(1)}'(0)=
\frac{d\eta_{(1)}}{d\lambda}(\lambda)\vert_{\lambda=0}=k_{(1)}^0(0)$})
and that of $\lambda_{(1)}$ and $\lambda_{(2)}$ with respect to
$\lambda_{(0)}$ (see equations (\ref{a74}) and (\ref{a75})), we
obtain \eq\label{backreacteddL} d_L\,=\,{\cal A}\,+\,{\cal
B}\,\widetilde{z}\,+\, {\cal C}\,\widetilde{z}^2\feq with
\begin{eqnarray}&& {\cal A}=\lambda_{(1)}(0) + \lambda_{(2)}(0) -
\lambda_{(1)}(0)\,\eta_{(1)}'(0)\;,\\
&& {\cal B}={\frac{1}{2}} + \lambda_{(1)}(0) -
{\frac{\theta_{(1)}(0,0)\,\lambda_{(1)}(0)}{4}} + \lambda_{(2)}(0)
- {\frac{\eta_{(1)}'(0)}{2}} -
\lambda_{(1)}(0)\,\eta_{(1)}'(0)\nonumber\\&&  -
{\frac{\eta_{(2)}'(0)}{2}} +{\frac{\lambda_{(1)}'(0)}{2}} -
{\frac{\eta_{(1)}'(0)\,\lambda_{(1)}'(0)}{2}}
+{\frac{\lambda_{(2)}'(0)}{2}} -
\lambda_{(1)}(0)\,\eta_{(1)}''(0)\;,\\
&& {\cal C}=\frac18\bigg(1 - \theta_{(1)}(0,0) -
{\theta_{(2)}(0,0)} -
  {\frac{\theta_{(1)}(0,0)\,\lambda_{(1)}(0)}{2}} - {\eta_{(1)}'(0)} +
  {\theta_{(1)}(0,0)\,\eta_{(1)}'(0)}\nonumber\\&& - {\eta_{(2)}'(0)}
  +
  {\lambda_{(1)}'(0)}- {\theta_{(1)}(0,0)\,\lambda_{(1)}'(0)} -
  {\eta_{(1)}'(0)\,\lambda_{(1)}'(0)} + {\lambda_{(2)}'(0)} - 2
\eta_{(1)}''(0) \nonumber\\&&-
 2{\lambda_{(1)}(0)\,\eta_{(1)}''(0)} - 4\lambda_{(1)}'(0)\,\eta_{(1)}''(0) -
 2{\eta_{(2)}''(0)} + {\lambda_{(1)}''(0)} -
 {\eta_{(1)}'(0)\,\lambda_{(1)}''(0)} + {\lambda_{(2)}''(0)}\nonumber\\&& -
  3\,\lambda_{(1)}(0)\,\eta_{(1)}'''(0) -
  2{\lambda_{(1)}(0)\,\theta_{(1)}^{\phantom{(1)}(0,1)}(0,0)} -
  {\frac{\lambda_{(1)}(0)\,\theta_{(1)}^{\phantom{(1)}(1,0)}(0,0)}{2}}\bigg)\;.
\end{eqnarray}
Using equations (\ref{a74}) and (\ref{a75}) together with
\eq\eta_{(1)}(0)=\eta_{(1)}'(0)=\eta_{(2)}(0)=\eta_{(2)}'(0)=\theta_{(1)}(0,0)=\theta_{(2)}(0,0)=0\feq
(the first four relations  are among the initial conditions we
have imposed on the geodesics equation, while the last two follow
from equation (\ref{serie})), we can easily get
\begin{eqnarray}&&{\cal A}=0\;,\\&& {\cal B}=\frac{1}{8}\,\left( 4 +
2{\eta_{(1)}''(0)}+ {\left(\eta_{(1)}''(0)\right)^2} +
2{\eta_{(2)}''(0)} \right)\;,\\&&{\cal C}=\frac{1}{32}\Big(4
-6\,\eta_{(1)}''(0) + 2\,\eta_{(1)}'''(0)
-4\,{{\eta_{(1)}''(0)}^2} - 6\,\eta_{(2)}''(0)\nonumber \\&&+
3\,{{ \eta }_{(1)}}''(0)\eta_{(1)}'''(0) + 2\,\eta_{(2)}'''(0)
\Big)\;.\end{eqnarray} Comparison between Eqs. (\ref{q0obsdef}) and
(\ref{backreacteddL}) then yields
\begin{equation}\label{prima} \widetilde{q}_0=\frac{1}{8}\Big[4 + 2\left( 4\eta_{(1)}''(0) - \eta_{(1)}'''(0) \right) +
{{\eta_{(1)}''(0)}^2} + 8\eta_{(2)}''(0) -
2{\eta}_{(1)}''(0)\eta_{(1)}'''(0) - 2\eta_{(2)}'''(0)
\Big]\;,\end{equation} \eq\label{seconda} \widetilde{{\cal H}}_0=2
- \eta_{(1)}''(0) - \eta_{(2)}''(0)\;.\feq We can express all the
parameters in terms of the original metric $\gamma_{ij}$
as\footnote{When applied to the metric, the prime stands for
differentiation with respect to conformal time:
$\phantom{k}^\prime=\partial_\eta$. }
\eq\eta_{(1)}''(0)=-\frac12\left(\gamma_{(1)ij}\right)^\prime
e^ie^j\;, \feq
\eq\eta_{(1)}'''(0)=\frac12\left(\gamma_{(1)ij}\right)^{\prime\prime}
e^ie^j -\frac12\left(\partial_k\gamma_{(1)ij}\right)^\prime
e^ie^je^k\;,\feq \eq
\eta_{(2)}''(0)=-\frac14\left(\gamma_{(2)ij}\right)^\prime
e^ie^j-\left(\gamma_{(1)ij}\right)^\prime e^i k^{(1)j}(0)\,
,\feq\begin{align}&\eta_{(2)}'''(0)=
\frac14\left(\gamma_{(2)ij}\right)^{\prime\prime}
e^ie^j-\frac14\left(\partial_k\gamma_{(2)ij}\right)^\prime
e^ie^je^k+\left(\gamma_{(1)ij}\right)^{\prime\prime} e^i
k_{(1)}^j(0)\nonumber\\&-e^k\partial_k\left(\gamma_{(1)ij}\right)^\prime
e^i k_{(1)}^j(0)-\left(\gamma_{(1)ij}\right)^\prime e^i
\frac{dk^{(1)j}}{d\lambda}(0)-\frac12\partial_l\left(\gamma_{(1)ij}\right)^\prime
e^ie^j k_{(1)}^l(0)\, ,\end{align} \eq
\frac{dk^i_{(1)}}{d\lambda}(0)=
-\partial_k\gamma^i_{(1)j}e^ke^j+e^k\left(\gamma^i_{(1)k}\right)^\prime
+\frac12\partial^i\gamma_{(1)kj}e^ke^j\;, \feq \eq
k^i_{(1)}(0)=-\frac12\gamma_{(1)kj}e^ke^je^i\;. \feq The functions
$\gamma_{(r)}$ ($r=1,2$) and their derivatives are evaluated at
the observer's position $\boldsymbol{x}_{(0)}=0$ at the present
time $\eta_0=1$. In terms of the gravitational potentials we have

\eq
\label{primamediafirst}\eta_{(1)}''(0)=\psi_{(1)}^\prime-\frac12\left(\chi_{(1)ij}\right)^\prime
e^ie^j%=\frac{1}{3}\varphi_{,ij}e^ie^j
\;,\feq
\eq\eta_{(1)}'''(0)=-\psi_{(1)}^{\prime\prime}+\frac12\left(\chi_{(1)ij}\right)^{\prime\prime}
e^ie^j
+e^k\partial_k\psi_{(1)}^\prime-\frac12e^k\left(\partial_k\chi_{(1)ij}\right)^\prime
e^ie^j
%=\frac{1}{3}\varphi_{,ijk}e^ie^je^k-\frac{1}{3}\varphi_{,ij}e^ie^j
\;,\feq
\begin{align}
&\eta_{(2)}''(0)=\frac12\psi_{(2)}^\prime-\frac14\left(\chi_{(2)ij}\right)^\prime
e^ie^j+2\psi_{(1)}^\prime e^i
k_{(1)i}(0)-\left(\chi_{(1)ij}\right)^\prime e^i
k^{(1)j}(0)%=\nonumber\\&-\frac{5}{18}\varphi^{,i}\varphi_{,i}+\frac{1}{42}\left(\varphi^{,ij}\varphi_{,ij}-\left(\nabla^2\varphi\right)^2\right)
%-\frac{19}{126}\varphi^{,i}_{\phantom{i},j}\varphi_{,ik}e^ke^j
%+\frac{5}{9}\left(\varphi_{,j}e^j\right)^2+\frac{2}{21}\nabla^2\varphi\varphi_{,ij}e^ie^j
%\nonumber\\&-\frac{1}{21}S^{jk}e_je_k-\frac{2}{3}\tau^{jk}e_je_k-\frac14\partial_\eta\pi^{ij}e_ie_j+\frac{2}{3}\varphi_{,ij}e^ik_{(1)}^j(0)
\;,\end{align}
\begin{align}
&\eta_{(2)}'''(0)=
-\frac12\psi_{(2)}^{\prime\prime}+\frac14\left(\chi_{(2)ij}\right)^{\prime\prime}
e^ie^j+\frac12e^k\partial_k\psi_{(2)}^\prime-\frac14e^k\left(\partial_k\chi_{(2)ij}\right)^\prime
e^ie^j-2\psi_{(1)}^{\prime\prime} e^i
k_{(1)i}(0)+\left(\chi_{(1)ij}\right)^{\prime\prime} e^i
k_{(1)}^j(0)\nonumber\\&+2e^j\partial_j\psi_{(1)}^\prime e^i
k_{(1)i}(0)-e^k\partial_k\left(\chi_{(1)ij}\right)^\prime e^i
k_{(1)}^j(0)+2\psi_{(1)}^\prime e^i
\frac{dk_{(1)i}}{d\lambda}(0)-\left(\chi_{(1)ij}\right)^\prime e^i
\frac{dk^{(1)j}}{d\lambda}(0)+\partial_i\psi_{(1)}^\prime
k_{(1)}^i(0)\nonumber\\&-\frac12\left(\partial_l\chi_{(1)ij}\right)^\prime
e^ie^j k_{(1)}^l(0)%=\nonumber\\&\frac{10}{9}e^{i}_{ }e^{j}_{
%}e^{k}_{ }{\varphi}_{,k}{\varphi}_{,ij} -\frac{5}{9}e^{j}_{
%}{\varphi}^{,k}{\varphi}_{,jk}  + \frac{2}{21}e^{i}_{ }e^{j}_{
%}e^{k}_{ }\nabla^2{\varphi}{\varphi}_{,ijk}- \frac{19}{63}e^{i}_{
%}e^{j}_{
%}e^{k}_{}{\varphi}^{,l}_{,\phantom{l}k}{\varphi}_{,ijl}+\frac{2}{21}
%e^{i}_{ }e^{j}_{
%}e^{k}_{}{\varphi}_{,jk}{\varphi}^{,l}_{,\phantom{l}li} + \frac{1}{21}
%e^{j}_{ }{\varphi}^{,kl}{\varphi}_{,jkl}\nonumber\\
%& - \frac{1}{21}e^{j}_{
%}\nabla^2{\varphi}{\varphi}^{,l}_{\phantom{l},lj}- \frac{1}{21}
%e^{i}e_{j}e_{k}S^{jk}_{\phantom{jk},i} - \frac{2}{3}
%e^{i}e_{j}e_{k}\tau^{jk}_{\phantom{jk},i}+\frac{5}{18}\varphi^{,i}\varphi_{,i}-\frac{1}{14}\left(\varphi^{,ij}\varphi_{,ij}-\left(\nabla^2\varphi\right)^2\right)
%+\frac{19}{42}\varphi^{,i}_{\phantom{i},j}\varphi_{,ik}e^ke^j\nonumber\\&-\frac{5}{9}\left(\varphi_{,j}e^j\right)^2-\frac{2}{7}\nabla^2\varphi\varphi_{,ij}e^ie^j+\frac{1}{7}S^{jk}e_je_k+\frac{2}{3}\tau^{jk}e_je_k-\frac14\partial_\eta\pi^{ij}_{\phantom{ij},k}e_ie_je^k+\frac14\partial^2_\eta\pi^{ij}e_ie_j
%-\frac{2}{3}\varphi_{,ij}e^ik_{(1)}^j(0)\nonumber\\
%&+\frac{2}{3}\varphi_{,ijk}e^ie^jk_{(1)}^k(0)
%+\frac{2}{3}\varphi_{,jk}e^j\frac{dk_{(1)}^k}{d\lambda}(0)+\frac{1}{3}e^ke^j\varphi_{,ikj}k^i_{(1)}(0)
\;,\end{align}
\begin{align}
&\frac{dk^i_{(1)}}{d\lambda}(0)=-\partial^i\psi_{(1)}+2
e^i\left(e^j\partial_j\psi_{(1)}\right)-2e^i\left(\psi_{(1)}\right)^\prime
-\partial_k\chi^i_{(1)j}e^ke^j+e^k\left(\chi^i_{(1)k}\right)^\prime
+\frac12\partial^i\chi_{(1)kj}e^ke^j%=\nonumber\\&-\frac{5}{3}\varphi^{,i}+\frac{10}{3}e^ie^j\varphi_{,j}+\frac{1}{6}e_je_k\varphi^{,ijk}-\frac{2}{3}e_j\varphi^{,ij}
\;,\end{align} \eq
k^i_{(1)}(0)=\left(\psi_{(1)}-\frac12\chi_{(1)kj}e^ke^j\right)e^i
\; ,
\label{primamedialast}\feq  where $\psi_{(r)}$, $\chi_{(r)}$ ($r=1,2$) %, $\varphi$, $\tau_{ij}$,
%$S_{ij}$, $\pi_{ij}$
and their derivatives are evaluated in the observer's position
$\boldsymbol{x}_{(0)}=0$ at the present time $\eta_0=1$.

What we need to make a comparison with the observations  are the
averages of these quantities over the sky (that is, over the
direction of observation).\\
Let us call $\textbf{m}$ the direction of observation (by
definition, $\textbf{m}\cdot\textbf{m}=1$): we can of course
expand $\textbf{m}$ either in the coordinate basis
$\left\{\frac{\partial}{\partial x^1},\frac{\partial}{\partial
x^2},\frac{\partial}{\partial x^3}\right\}$ or in a Cartesian
system of axes $\left\{n_{(1)},n_{(2)},n_{(3)}\right\}$ (a
''triad'' carried by the observer): \eq\label{cosines1}
\textbf{m}=m^i\frac{\partial}{\partial x^i}=\epsilon^i n_{(i)}\;.
\feq In our case, the $e^i$, $i=1,2,3$, are proportional to the
$m^i$, $i=1,2,3$: \eq\label{cosines2} e^i=N\, m^i\;,\feq where the
constant $N$ is determined by the normalization condition Eq.
(\ref{a32}). In order to perform an angular average, we therefore
need  to express equations
(\ref{primamediafirst})-(\ref{primamedialast}) not in terms of the
$e^i$, but in terms of the director cosines $\epsilon^i$.\\
Let us therefore define \eq\label{cosines3}
n_{(k)}=(1+\psi_{(1)})\frac{\partial}{\partial
x^k}-\frac12\chi_{(1)k}^{i}\frac{\partial}{\partial x^i}\;; \feq
to first order we have \eq n_{(i)}\cdot n_{(j)}=\delta_{ij}\;:\feq
the
$n_{(i)}$, $i=1,2,3$, are the unit vectors of a triad.\\
Taking then into account Eqs. (\ref{a32}), (\ref{cosines1}),
(\ref{cosines2}) and (\ref{cosines3}) we get, to first order,
\begin{multline}
e^i=\sqrt{g_{kl}e^ke^l}\frac{e^i}{\sqrt{g_{kl}e^ke^l}}=\sqrt{g_{kl}e^ke^l}\,m^i=\\\left(1-\psi_{(1)}+\frac12\chi_{(1)}^{kl}e_ke_l\right)
\left[(1+\psi_{(1)})\epsilon^i-\frac12\chi_{(1)}^{ij}\epsilon_j\right]=
\epsilon^i-\frac12\chi_{(1)}^{ij}\epsilon_j+\frac12\epsilon^i\left(\chi_{(1)}^{kl}\epsilon_k\epsilon_l\right)
\;.\end{multline} Performing this substitution in equations
(\ref{primamediafirst})-(\ref{primamedialast}), we can calculate
the angular averages of $\widetilde{{\cal H}}_0$ and
$\widetilde{q}_0$ by means of the following identities (which can
be obtained by expressing the director cosines $\epsilon^i$ in
spherical coordinates): \eq\langle
\epsilon^i\rangle_\Omega=\langle
\epsilon^i\epsilon^j\epsilon^k\rangle_\Omega=\langle
\underbrace{\epsilon^i\epsilon^j\ldots \epsilon^k}_{2n+1\,\rm
terms}\rangle_\Omega=0\;,\feq \eq\langle
\epsilon^i\epsilon^j\rangle_\Omega=\frac13\delta^{ij}\;,\feq
\eq\langle
\epsilon^i\epsilon^j\epsilon^k\epsilon^l\rangle_\Omega=\frac{1}{15}\left(\delta^{ij}\delta^{kl}+\delta^{ik}\delta^{jl}+\delta^{il}\delta^{kj}\right)\;,
\feq where we have indicated with
$\langle\phantom{e^i}\rangle_\Omega$ the average over solid angle
\eq\langle\phantom{e^i}\rangle_\Omega:=\frac{1}{4\pi}\int
d\Omega\;.\feq

Therefore, by using these identities and equations
(\ref{a40})-(\ref{a46}), the angular averages of equations
(\ref{prima}) and (\ref{seconda}) become, after resurrecting the
dependence on $\eta_0$,
\begin{align}\label{H0final}
\langle\widetilde{\cal H}_0\rangle_\Omega&={\cal H}_{0}
\Bigg[1-\eta_0\left(\frac12\psi_{(1)}^\prime+\frac14\psi_{(2)}^\prime
+\psi_{(1)}\psi_{(1)}^\prime+\frac{1}{12}\left(\chi_{(1)}^{ij}\right)^\prime\chi_{(1)ij}\right)\Bigg]\nonumber\\&={\cal
H}_{0}\bigg[1-\left(\frac{1}{18}\nabla^2\varphi-\frac{5}{108}\left(\nabla\varphi\right)^2+\frac{5}{27}\varphi\nabla^2\varphi\right)\eta_0^2
-\left(\frac{1}{189}\varphi^{,ij}\varphi_{,ij}+\frac{1}{252}\left(\nabla^2\varphi\right)^2\right)\eta_0^4\,\bigg]\;,
\end{align} \begin{align}\label{q0final}
\langle\tilde{q}_0\rangle_\Omega=&q_{0}\Bigg[1
+\eta_0\Bigg(2\psi_{(1)}^\prime+\psi_{(2)}^\prime+
4\psi_{(1)}\psi_{(1)}^\prime+\frac{1}{3}\left(\chi_{(1)}^{ij}\right)^{\prime}\chi_{(1)ij}\Bigg)\nonumber\\&
+\eta_0^2\Bigg(\frac12\psi_{(1)}^{\prime\prime}
+\frac14\psi_{(2)}^{\prime\prime}+\frac94\left(\psi_{(1)}^\prime\right)^2
+\psi_{(1)}\psi_{(1)}^{\prime\prime}+\frac{7}{40}\left(\chi_{(1)ij}\right)^\prime\left(\chi_{(1)}^{ij}\right)^\prime
+\frac1{12}\left(\chi_{(1)}^{ij}\right)^{\prime\prime}\chi_{(1)ij}\Bigg)\nonumber\\&
+\eta_0^3\Bigg(\frac12\psi_{(1)}^\prime\psi_{(1)}^{\prime\prime}
+\frac{1}{60}\left(\chi_{(1)ij}\right)^\prime\left(\chi_{(1)}^{ij}\right)^{\prime\prime}\Bigg)
\Bigg]\nonumber\\=&q_{0}\bigg[1+\left(\frac{5}{18}\nabla^2\varphi+\frac{25}{27}\varphi\nabla^2\varphi-\frac{25}{108}\left(\nabla\varphi\right)^2\right)\eta_0^2+
\left(\frac{23}{270}\varphi^{,ij}\varphi_{,ij}+\frac{1}{30}\left(\nabla^2\varphi\right)^2\right)\eta_0^4\bigg]\;,\end{align}
where $\psi_{(r)}$, $\chi_{(r)}$ ($r=1,2$) and $\varphi$ are
evaluated at the observer's position $\boldsymbol{x}_{(0)}=0$ and
at the present time $\eta_0$ and where ${\cal H}_0$ and $q_0$ are
the background Hubble constant and deceleration parameter ${\cal
H}_0=\frac{2}{\eta_0}$ and $q_0=\frac12$.

\end{widetext}

\end{document}